\documentclass[prd,aps,twocolumn,a4paper,floatfix,10pt]{revtex4-1}

\usepackage[utf8]{inputenc}
\usepackage{amsmath, amsthm, amsfonts, amssymb, latexsym}
\usepackage{graphicx}
\usepackage{color}
\usepackage{subfigure}
\usepackage{url}
\usepackage{psfrag}
\usepackage{mathrsfs}
\usepackage{multirow}
\usepackage{comment}
\usepackage[normalem]{ulem}
\usepackage{tikz} 
\usepackage{bbold}
\usepackage{bbm}
\usepackage{flushend}

\def\p{\partial}
\DeclareMathAlphabet\mathbfcal{OMS}{cmsy}{b}{n}
\newcommand{\s}{{\mathbbmss{s}}}

\usepackage[unicode]{hyperref}
\hypersetup{
    unicode=true,          
    pdftitle={Hyperbolicity of General Relativity in Bondi-like gauges},
    pdfauthor={},
    pdfsubject={},
    pdfcreator={LaTeX},          
    pdfproducer={LaTeX},         
    pdfkeywords={},
    colorlinks=false,            
  }

\DeclareSymbolFont{matha}{OML}{txmi}{m}{it}
\DeclareMathSymbol{\varv}{\mathord}{matha}{118}

\graphicspath{ {./images/} }

\begin{document}

\title{
Hyperbolicity of General Relativity in Bondi-like gauges}

\author{Thanasis Giannakopoulos}
\author{David Hilditch}
\author{Miguel Zilh\~ao}

\affiliation{
  Centro de Astrof\'{\i}sica e Gravita\c c\~ao -- CENTRA,
  Departamento de F\'{\i}sica, Instituto Superior T\'ecnico -- IST,
  Universidade de Lisboa -- UL,
  Av.\ Rovisco Pais 1, 1049-001 Lisboa, Portugal
}

\begin{abstract}
Bondi-like (single-null) characteristic formulations of general
relativity are used for numerical work in both asymptotically flat and
anti-de Sitter spacetimes. Well-posedness of the resulting systems of
partial differential equations, however, remains an open question.
The answer to this question affects accuracy, and potentially the
reliability of conclusions drawn from numerical studies based on such
formulations. A numerical approximation can converge to the continuum
limit only for well-posed systems; for the initial value problem in
the~$L^2$ norm this is characterized by strong hyperbolicity. We find
that, due to a shared pathological structure, the systems arising from
the aforementioned formulations are however only weakly hyperbolic. We
present numerical tests for toy models that demonstrate the
consequence of this shortcoming in practice for the characteristic
initial boundary value problem. Working with alternative norms in
which our model problems may be well-posed we show that convergence
may be recovered. Finally we examine well-posedness of a model for
Cauchy-Characteristic-Matching in which model symmetric and weakly
hyperbolic systems communicate through an interface, with the latter
playing the role of GR in Bondi gauge on characteristic slices. We
find that, due to the incompatibility of the norms associated with the
two systems, the composite problem does not naturally admit energy
estimates.
\end{abstract}

\maketitle

\section{Introduction} \label{Section:Introduction}

\begin{figure}[!t]
  \includegraphics[width=0.16\textwidth]{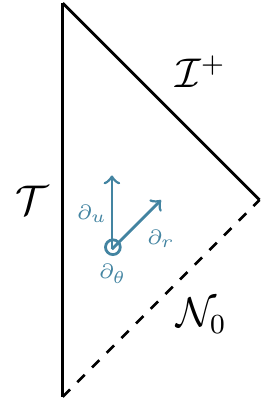}
  \caption{The CIBVP for the wave-zone of an asymptotically flat
    spacetime. Boundary data are given on the timelike inner
    boundary~$\mathcal{T}$, the worldtube~$r=r_0$, and initial data on
    the null hypersurface~$\mathcal{N}_0$ of constant retarded
    time~$u_0$.}
  \label{flat_CIBVP}
\end{figure}

Characteristic formulations of General Relativity (GR) have advantages
over more standard spacelike foliations in a number of situations. For
instance, in the asymptotically flat setting, the Bondi-Sachs
formalism~\cite{BonBurMet62,Sac62}, crucial to the modern
understanding of gravitational waves, underpins codes that aim to
produce waveforms of high accuracy. This approach exploits the fact
that null hypersurfaces reach future null infinity and hence allows
the avoidance of systematic errors from extrapolation techniques. The
general setup in these approaches is to construct a standard Cauchy
problem in a finite region of the spacetime, where the main action,
such as the collision of two black holes, takes place. The data on the
worldtube of this finite region serve as boundary data for the
characteristic initial boundary value problem (CIBVP). Solving this
CIBVP one can compute quantities such as the gravitational wave news
function at future null infinity. This method is often called
\textit{Cauchy-characteristic extraction} (CCE)~\cite{BisGomLeh97a,
  BisGomLeh97, ZloGomHus03, HanSzi14, BarMoxSch19, MoxSchTeu20}. If
the Cauchy and the CIBVP are solved simultaneously and one attempts to
match the worldtube data from both the Cauchy problem and the CIBVP,
then the method is called \textit{Cauchy-characteristic matching}
(CCM), see~\cite{Win12, Szi00} for a thorough review. In
Fig.~\ref{flat_CIBVP} an illustration of the geometric setup is
given. Concerning asymptotically anti-de Sitter (AdS) spacetimes,
characteristic formulations of GR are widely used in the field of
numerical holography, which provides insights into the behavior of
strongly coupled matter~\cite{CheYaf11, AttCasMat17}. We refer to the
aforementioned characteristic formulations as \textit{Bondi-like} or
\textit{single-null}.

A practical advantage of Bondi-like gauges is that the field equations
can then be written as a set of nested differential equations which
can be efficiently solved. For the resulting CIBVP one provides data
on a timelike boundary and initial data on either an outgoing or
ingoing null hypersurface depending on the physical setup. There are
many examples of numerical codes making successful use of this
formalism. Since these codes have successfully passed a multitude of
convergence tests, and in various physical contexts, one might say
that there is numerical evidence that the PDE problem solved is
well-posed. To the best of our knowledge however a proof of this
result is missing. By well-posedness we mean the usual notion that the
problem admits unique solutions that depend continuously on the given
data in a suitable norm. Interest in this property is not purely
mathematical, since a numerical solution can converge to the
continuous one only for well-posed PDE problems. The PDE systems that
interest us here are of the hyperbolic class. A necessary condition
for well-posedness of these systems in~$L^2$, or in fact suitable
Sobolev norms, is that they are strongly
hyperbolic~\cite{GusKreOli95,Hil13}. Specifically, we consider PDEs in
the generic form
\begin{align}
  \mathbfcal{A}^t(\mathbf{u},x^\mu) \, \p_t \mathbf{u}
  + \mathbfcal{A}^p(\mathbf{u},x^\mu)\,
  \p_p \mathbf{u} +  \mathbfcal{S}(\mathbf{u},x^\mu) = 0
  \,,\label{gen_hyper_PDE}
\end{align}
where~$\mathbf{u} = (u_1, u_2, \dots, u_q )^T \, ,$ is the state
vector of the system and
\begin{align*}
  \mathbfcal{A}^\mu =
  \begin{pmatrix}
    a^{\mu}_{11}  & \dots  & a^{\mu}_{1q} \\
    \vdots & \ddots & \vdots \\
    a^{\mu}_{q1}  & \dots  & a^{\mu}_{qq} 
  \end{pmatrix}
\end{align*}
denotes the principal part matrices, with~$ \det(\mathbfcal{A}^t) \neq
0 \, $. To classify locally the character of the PDE we linearize
about a background solution and then work pointwise in the frozen
coefficient approximation, henceforth suppressing the explicit
dependencies of the principal part matrices and source vector, and
requesting the following definitions everywhere. We can construct the
principal symbol
\begin{align}
  \mathbf{P}^s = \left(\mathbfcal{A}^t \right)^{-1}
  \mathbfcal{A}^p \, s_p \,,\label{principal_symbol}
\end{align}
where~$s^i$ is an arbitrary unit spatial vector. If~$\mathbf{P}^s$ has
real eigenvalues for all~$s^i$, then the PDE system is called
\textit{weakly hyperbolic} (WH), whereas if in addition~$\mathbf{P}^s$
is diagonalizable for all~$s^i$, and there exists a constant~$K$
independent of~$s^i$ such that
\begin{align*}
|\mathbf{T}_s|+|\mathbf{T}_s^{-1}|\leq K,
\end{align*}
with~$\mathbf{T}_s$ the similarity matrix that
diagonalizes~$\mathbf{P}^s$, it is called \textit{strongly
  hyperbolic}~(SH).

Presently we analyze the character of the PDE systems that arise in
two specific Bondi-like formulations of GR. The original systems
involve second order derivatives, so we perform reductions to first
order to conveniently build the principal parts. We find that, due to
a degeneracy in the angular/transverse principal parts, these
formulations are only WH. Consequently, they give rise to PDE problems
that are ill-posed in~$L^2$ even in the linear, frozen coefficient
approximation, which prohibits well-posedness of the full system in
associated Sobolev norms. We argue furthermore that this result holds
true for every possible first order reduction.

Subsequently we perform careful numerical experiments that demonstrate
the consequence of this shortcoming in practice. We work with two toy
models, one of which is SH and the other only WH. We perform
robust-stability-like~\cite{BabHusAli08,applesweb1} tests, suitably
modified for the characteristic setting, and find that convergence in
a discrete approximation to~$L^2$ is prohibited in the WH
model. Convergence with the latter model can be achieved by using a
discrete approximation to a modified norm that involves a subset of
derivatives of the state vector fields and adjusting the initial data
for the test.

The structure of the paper is as follows. In
Sec.~\ref{Section:characteristic_formulations} we give an overview of popular
Bondi-like formulations of GR in both the asymptotically flat and AdS
contexts, and present our hyperbolicity analysis of each. Afterwards,
in Sec.~\ref{Section:toy_models} we present our toy models, then in
Sec.~\ref{Section:numerical_experiments} we present numerical experiments
demonstrating the effect of our analytic results in practice. Finally
we conclude in Sec.~\ref{Section:conclusions}. Geometric units are used
throughout.

\section{Characteristic formulations}
\label{Section:characteristic_formulations}

In this section we present two characteristic formulations of GR in
Bondi-like gauges that are widely used in numerical work. The first,
the Bondi-Sachs formulation proper, is popular in the asymptotically
flat setting, whereas the second, known as the affine-null system, is
used most often in numerical holography. We demonstrate that each is
only weakly hyperbolic.

\subsection{Bondi-Sachs Gauge} \label{Subsection:BS_gauge}

In Bondi-Sachs gauge~\cite{BonBurMet62,Sac62} a generic 4-dimensional
axially symmetric metric can be written as
\begin{align}
  ds^2 &= \left( \frac{V}{r} e^{2 \beta}
  - U^2 r^2 e^{2\gamma} \right) \, du^2 
  + 2 e^{2 \beta} du \, dr   \label{BS_metric_ansatz}\\
& + 2 U r^2 e^{2 \gamma} \, du \, d\theta
  - r^2 \left( e^{2 \gamma} \, d\theta^2
  + e^{-2 \gamma} \sin^2\theta \, d\phi^2 \right) \,.\nonumber
\end{align}
Here~$u$ is a null coordinate, called retarded time, $r$ is the areal
radius, and~$\theta,\phi$ give coordinates on the two-sphere in the
standard way. All metric functions are functions of~$(u,
r,\theta)$. To make contact with~\cite{Win12} we adopt the signature
convention~$(+,-,-,-)$. In this formulation Einstein's equations
exhibit a nested structure. For axially symmetric spacetimes the PDE
system consists of three equations intrinsic to the hypersurfaces of
constant time,
\begin{equation}
\begin{aligned}
  &\beta_{,r} = \frac{1}{2} r \left( \gamma_{,r} \right)^2 \, ,\\
  &\left[r^4 e^{2 (\gamma-\beta)} U_{,r} \right]_{,r}
   = \\
  &\qquad
   2r^2\left[r^2 \left( \frac{\beta}{r^2} \right)_{,r\theta}
   - \frac{\left(\sin^2 \theta \, \gamma \right)_{,r \theta}}
   {\sin^2 \theta}
   + 2 \gamma_{,r} \, \gamma_{,\theta} \right] \,,\\
  &V_{,r} = - \frac{1}{4} r^4 e^{2(\gamma-\beta)} \left(U_{,r} \right)^2
   + \frac{\left( r^4 \,
     \sin \theta \,  U\right)_{,r \theta}}{2 r^2 \sin \theta}\\
  & \quad\quad
  + e^{2(\beta-\gamma)} \Big[ 1 - \frac{\left( \sin\theta \, \beta_{,\theta}
    \right)_{, \theta}}{\sin \theta} + \gamma_{, \theta \theta}
    + 3 \cot \theta \, \gamma_{,\theta} \\
  & \qquad \qquad \qquad \qquad 
    - \left( \beta_{,\theta} \right)^2 - 2 \gamma_{,\theta}
    \left(\gamma_{,\theta} - \beta_{, \theta} \right)  \Big]\,,
  \end{aligned}
\end{equation}
and one equation that involves extrinsic derivatives,
\begin{align}
  &4r \left( r \gamma \right)_{,u r}
  = \left\{ 2 r \, \gamma_{,r} \, V - r^2 \left[ 2 \gamma_{,\theta} \, U
    + \sin\theta \left( \frac{U}{\sin \theta}
    \right)_{,\theta} \right] \right\}_{,r}
  \nonumber\\
  & \qquad \qquad
  - 2r^2 \frac{\left(\gamma_{,r} \, U \, \sin \theta
    \right)_{,\theta}}{\sin \theta}
  + \frac{1}{2} r^4 e^{2(\gamma-\beta)} \left(U_{,r} \right)^2
  \nonumber \\
  & \qquad \qquad
  + 2 e^{2(\beta-\gamma)} \left[ \left( \beta_{,\theta} \right)^2
    + \sin \theta \left( \frac{\beta_{,\theta}}{\sin \theta}
    \right)_{,\theta} \right]\, .
  \label{evol_eq}
\end{align}
The remaining Einstein equations are not solved explicitly and, as in
any other free-evolution approach, are therefore ignored in our
analysis.

\subsubsection{First order reduction \& Linearization}

In~\cite{FriLeh99} and~\cite{GomFri03} the authors studied existence
and uniqueness of the CIBVP for the formulation given in the previous
subsection. They considered the linearized and quasilinear systems,
but did not study continuous dependence on given data, which will be
our main focus. To treat the system in the original higher-order
derivative form, we could follow~\cite{GunGar05,HilRic13a}. But for
convenience in building the principal parts we instead perform an
explicit first order reduction. Since this PDE is built as a
reduction, there is the subtlety of the associated constraints and the
specific choice of reduction, which we discuss in detail later. The
minimal set of reduction variables are given by
\begin{align*}
  U_r = \p_r U \,,
  \gamma_r = \p_r \gamma \,,
  \gamma_\theta = \p_\theta \gamma \,,
  \beta_\theta = \p_\theta \beta\,.
\end{align*}
We linearize the resulting equations about a fixed
background. In~\cite{GiaHilZil20_public} one can find the complete
analysis for both Minkowski and arbitrary backgrounds. The resulting
level of hyperbolicity of the system is the same regardless, and so we
present the former for brevity. After this procedure the system reads
\begin{equation}
\begin{aligned}
  &\p_r \beta = 0 \, , \\
  &\p_r U_r  - \frac{2}{r^2}
   \p_r\beta_\theta + \frac{2}{r^2}
   \p_r\gamma_\theta + S_2 = 0 \,, \\
  &\p_r V +  \p_\theta \beta_\theta
   - \p_\theta \gamma_\theta
    - 2 r \p_\theta U - \frac{r^2}{2}
    \p_\theta U_r + S_3 = 0 \,,\\
  & 4 r^2 \p_u \gamma_r + 4 r \p_u \gamma -2 r^2 \,
    \p_r \gamma_r  \\
  & \qquad\qquad\;\;
    + 2 r \, \p_\theta U
    + r^2 \p_\theta U_r - 2 \p_\theta
    \beta_\theta + S_4 = 0 \, ,\\
  & \p_rU + S_5 = 0 \,,\\
  & \p_r \gamma + S_6 = 0 \,,\\
  & \p_r \gamma_\theta - \p_\theta \gamma_r = 0 \,,\\
  & \p_r \beta_\theta = 0 \,,
\end{aligned}
  \label{BS_linear_pde}
\end{equation}
where~$S_i$ denotes the various source terms and we work in the frozen
coefficient approximation, so that~$r$ and so forth must be treated as
constants. The variables can be collected in the state vector
\begin{align*}
  \mathbf{u} = \left( \beta\, , \gamma \, , U \, , V \, ,\gamma_r \, ,
  U_r \, , \beta_\theta \, ,\gamma_\theta  \right)^T\,,
\end{align*}
and the system can be written in the form~\eqref{gen_hyper_PDE} with
\begin{align} 
  \mathbfcal{A}^u \p_u \mathbf{u} + \mathbfcal{A}^r \p_r \mathbf{u}
  + \mathbfcal{A}^\theta \p_\theta \mathbf{u} + \mathbfcal{S} = 0.
  \label{flat_system_matrix_form_1}
\end{align}
The principal part matrix~$\mathbfcal{A}^u$ associated with retarded
time~$u$ is not invertible (see~\cite{GiaHilZil20_public} for the full
calculation). In order to use the standard definitions given in the
introduction we need a principal part associated to time derivatives
that is invertible. We achieve this by performing a coordinate
transformation to a frame that involves one timelike and three
spacelike directions.

\subsubsection{Coordinate transformation}

\begin{figure}[!t]
  \includegraphics[width=0.46\textwidth]{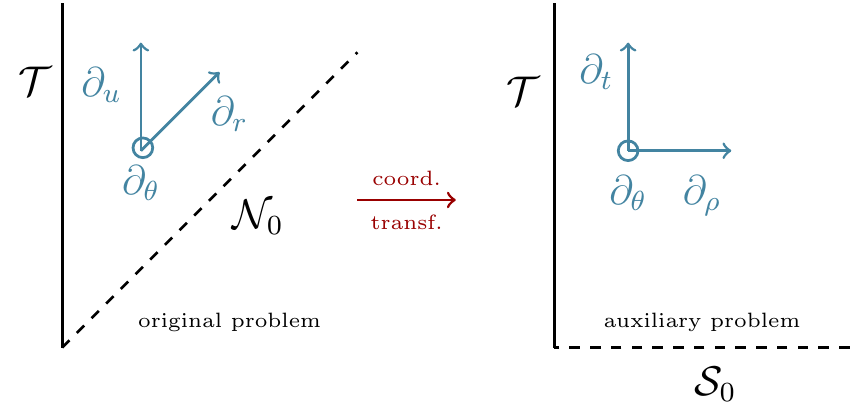}
  \caption{The original CIBVP is transformed into an auxiliary frame
    using the coordinate transformation~\eqref{flat_coord_transf} as
    shown. This allows the use of textbook definitions of
    hyperbolicity but does not affect the solution space.
  \label{flat_CIBVP_to_IBVP}}
\end{figure}

We wish to bring the system~\eqref{flat_system_matrix_form_1} to the
form~\eqref{gen_hyper_PDE}, which has a trivial time principal part
matrix
\begin{align*}
  \p_t \mathbf{u} + \mathbfcal{A}^p \p_p \mathbf{u}
  + \mathbfcal{S} =0\,,
\end{align*}
where~$\p_p$ denotes spatial derivatives, $\mathbf{u}$ denotes the
state vector. We therefore perform the following concrete coordinate
transformation
\begin{align}
  u = t-\rho \, , \qquad r=\rho \,,
  \label{flat_coord_transf}
\end{align}
with the angular coordinates unchanged, which yields the following
relation between the old and new basis vectors,
\begin{align*}
  \p_u = \p_t \, , \qquad \p_r  = \p_t
  + \p_\rho  \,,
\end{align*}
with the remaining vectors unaltered. A schematic of the auxiliary
setup is given in Fig.~\ref{flat_CIBVP_to_IBVP}. Applying the
transformation yields
\begin{align*}  
  \mathbfcal{A}^t \, \p_t  \mathbf{u} 
  + \mathbfcal{A}^r \, \p_\rho  \mathbf{u}
  + \mathbfcal{A}^\theta \, \p_\theta \mathbf{u} \, + \mathbfcal{S} = 0
  \, ,
\end{align*}
with~$\mathbfcal{A}^t = \mathbfcal{A}^u + \mathbfcal{A}^r$
invertible. After multiplying on the left with the inverse
of~$\mathbfcal{A}^t$ we bring the system to the desired form,
\begin{align}
  \p_t \mathbf{u} 
  + \mathbf{B}^\rho \, \p_\rho  \mathbf{u}
  + \mathbf{B}^{\hat{\theta}} \, \p_{\hat{\theta}}  \mathbf{u} \, + \mathbf{S} = 0
  \, , \label{flat_new_frame_system_2}
\end{align}
where~$\mathbf{B}^\rho = \left( \mathbfcal{A}^t \right)^{-1} \,
\mathbfcal{A}^r$ and~$\mathbf{B}^{\hat{\theta}} = \rho \left(
\mathbfcal{A}^t \right)^{-1} \, \mathbfcal{A}^\theta$
with~$\p_{\hat{\theta}} \equiv 1/\rho\,\p_\theta \, $,
and~$\mathbf{S}$ was redefined in the obvious manner. The solution
space in this frame is equivalent to that of the original one, so in
this sense the character of the PDE is invariant. For our system, the
principal part matrix~$\mathbf{B}^\rho$ is diagonalizable with real
eigenvalues.  Although~$\mathbf{B}^{\hat{\theta}}$ has real
eigenvalues, it does not have a complete set of eigenvectors, and
hence is not diagonalizable. Therefore the system resulting from the
specific first order reduction we made is only WH. In~\cite{Fri04} a
subsystem of a similar first-order reduction was shown to be symmetric
hyperbolic. Here, however, we are concerned with the best estimates
that can be made for the full system. In Sec.~\ref{Section:toy_models}
this is written up explicitly for our homogeneous WH model.

So far we have not ruled out the existence of an alternative first
order reduction that {\it is} SH however. To examine this possibility
we have to understand if any potential addition of reduction
constraints can make the system SH. The reduction constraints are
\begin{align}
  \p_\theta \beta - \beta_\theta
  &  = 0 \, ,\qquad
  \p_\theta \gamma - \gamma_\theta
  = 0 \, ,
    \label{flat_red_constr}
\end{align}
The definitions of the variables~$\gamma_r$ and~$U_r$ are solved
explicitly as time evolution equations within the
system~\eqref{BS_linear_pde} and therefore do not have an associated
constraint. This subtlety, along with an examination of the form of
the degeneracy follows in the next section.

\subsubsection{Generalized characteristic variables}
\label{Subsubsection:BS_gauge:GCV}

To understand the nature of the degeneracy
of~$\mathbf{B}^{\hat{\theta}}$ physically it is useful to consider the
generalized eigenvalue problem,
\begin{align*}
  \mathbf{l}_{\lambda_i} \left( \mathbf{B}^{\hat{\theta}}
  - \lambda_i \mathbb{1} \right)^m =0\,,
\end{align*}
with~$\lambda_i$ standing for the various eigenvalues,
and~$\mathbf{l}_{\lambda_i}$ representing either a true eigenvector
when~$m=1$ or else a generalized eigenvector when~$m>1$. The
eigenvalues of~$\mathbf{B}^{\hat{\theta}}$ are~$\lambda = \pm 1$, each
with algebraic multiplicity one and~$\lambda=0$ with algebraic
multiplicity six. The geometric multiplicity of each of~$\lambda = \pm
1/\rho$ is also one, but~$\lambda=0$ has geometric multiplicity five.
In other words one associated eigenvector is missing and we obtain one
nontrivial generalized eigenvector with~$m=2$
for~$\lambda=0$. Defining the invertible
matrix~$\mathbf{T}_{\hat{\theta}}^{-1}$ with the
vectors~$\mathbf{l}_{\lambda_i}$, as rows, we obtain the Jordan normal
form of the principal symbol in the~$\theta$ direction by the
similarity transformation
\begin{align*}
  \mathbf{J}^{\hat{\theta}} \equiv \mathbf{T}_{\hat{\theta}}^{-1} \,
  \mathbf{B}^{\hat{\theta}} \, \mathbf{T}_{\hat{\theta}}\,.
\end{align*}
The same matrix can be used to construct the generalized
characteristic variables of the system in the~$\theta$ direction,
namely the components
of~$\mathbf{v}\equiv \mathbf{T}_{\hat{\theta}}^{-1} \,
\mathbf{u}$. These are of course nothing more than the left
generalized eigenvectors contracted with the state vector. Working as
before in the frozen coefficient approximation, focusing on
the~$t,\theta$ parts of~\eqref{flat_new_frame_system_2}, and
multiplying on the left with~$\mathbf{T}_{\hat{\theta}}^{-1}$ we get
\begin{align}
  \p_t \mathbf{v} + \mathbf{J}_{\hat{\theta}}  \, \p_{\hat{\theta}} \mathbf{v}
       \simeq 0 \, ,\label{gen_char_system}
\end{align}
with~$\simeq$ denoting here equality up to non-principal terms and
spatial derivatives transverse to~$\p_{\hat{\theta}}$. The generalized
characteristic variables with speed (eigenvalue) zero are
\begin{align*}
  &\rho \, U + \frac{\rho^2}{2} \,U_r -\beta_\theta
  + \gamma_\theta \,,&\quad
  &\beta_\theta \,,&\quad &V\,\\
  & \rho \left(
    -2\rho U - \frac{\rho^2}{2} U_r
    + \beta_\theta - \gamma_\theta
    \right) \,,&\quad
  &\gamma\,,
  &\quad
  &\beta \,,
\end{align*}
of which the third and fourth are associated with the
non-trivial~$2\times2$ Jordan block
within~$\mathbf{J}^{\hat{\theta}}$.  Likewise we have
\begin{align*}
  -\frac{\rho}{2}U + \frac{\rho}{2}\gamma_r
  - \frac{\rho^2}{4} U_r + \frac{1}{2} \beta_\theta \,,\\
  -\frac{\rho}{2}U - \frac{\rho}{2}\gamma_r
  - \frac{\rho^2}{4} U_r + \frac{1}{2} \beta_\theta \,,
\end{align*}
with speeds~$\pm 1$ respectively. The structure
of~$\mathbf{J}_{\hat{\theta}}$ thus yields
\begin{equation}
\begin{aligned}
&- \p_t\left(2\rho U +\frac{\rho^2}{2} U_r
      - \beta_\theta + \gamma_\theta\right) \simeq 0\,,\\
&\p_tV - \rho \, \p_{\hat{\theta}}\left(2\rho U +\frac{\rho^2}{2} U_r
  - \beta_\theta + \gamma_\theta\right) \simeq 0 \,.
\end{aligned}
\label{K3_gen_char_var}
\end{equation}
Strongly hyperbolic systems admit a complete set of characteristic
variables in each direction. In other words, if our system were
strongly hyperbolic then up to non-principal and transverse derivative
terms each component of~$\mathbf{v}$ would satisfy an advection
equation. Presently the best we can achieve for~$V$ however
is~\eqref{K3_gen_char_var}. Physically we may therefore understand
weak hyperbolicity as the failure of~$V$, a {\it generalized}
characteristic variable, to satisfy such an advection equation. As
mentioned in the previous section, we could try and cure the equations
by using a different first order reduction. Observe that the choice of
different reductions corresponds to the freedom to add (derivatives
of) the reduction constraints to~\eqref{K3_gen_char_var} without
introducing second derivatives. As~$V$ appears at most once
differentiated in the original equations there is no associated
constraint, so we must hope to eradicate the~$\p_\theta$ term
from~\eqref{K3_gen_char_var} using~\eqref{flat_red_constr} without
introducing second derivatives. Even if the
variable~$U_\theta=\p_\theta U$ were introduced in the reduction
however, the~$\p_\theta \beta_\theta$ and~$\p_\theta\gamma_\theta$
terms would obviously persist. Thus one non-trivial {\it generalized}
characteristic variable always survives and prevents the existence of
a complete set of characteristic variables. Hence within the
coordinate basis built from~$(t,\rho,\theta)$, the field equations are
at best only weakly hyperbolic regardless of the specific reduction.

\subsection{Affine-null gauge} \label{Subsection:AN_gauge}

Although sometimes used in the asymptotically flat
setting~\cite{Win13,CreOliWin19}, the affine-null gauge is
particularly popular for evolutions in asymptotically AdS
spacetimes~\cite{CheYaf13}. For concreteness we will treat the
specific system that occurs in the case of asymptotically~$AdS_5$
spaces with planar symmetry, but we expect similar results in other
contexts with analogous gauges. The metric is written as
\begin{align}
  ds^2 &= -A d\varv^2 + \Sigma^2
  \left[ e^B dx_\perp^2 + e^{-2B} dz^2 \right]\nonumber\\
  &\quad + 2dR \,d\varv + 2 F d\varv dz\, . \label{EF_metric ansatz}
\end{align}
Here~$\varv$ denotes a null coordinate, called advanced time, $R$ is
called the holographic coordinate, and increases from the bulk of the
spacetime towards the boundary. All metric components are functions
of~$(\varv, R, z)$. We also denote by~$dx_\perp^2$ the flat metric in
the plane spanned by~$x_\perp$, the two coordinates associated with
the symmetry. Using the convenient definitions
\begin{equation}
\begin{aligned}
  d_z &\equiv \p_z -F \p_R \, ,\\
  d_+ &\equiv \p_\varv + \tfrac{A}{2} \p_R \,,
\end{aligned}
\label{tilde_dot_def}
\end{equation}
the field equations can be succinctly stated, and are
\begin{equation}
\begin{aligned}
 & \p_R^2\Sigma
  = - \frac{1}{2} \left( \p_R B \right)^2 \Sigma
   \, , \\
 &\Sigma^2 \, \p_R^2 F
 = \Sigma \left( 6 \, d_z \Sigma \, \p_R B + 4 \, \p_R d_z \Sigma
 + 3 \, \p_R F\, \p_R \Sigma \right)
    \\
 & \qquad
   + \Sigma^2 \left( 3 \, d_z B \, \p_R B + 2\, \p_R d_z B\right)
   - 4\, d_z \Sigma \, \p_R \Sigma
   \, , \\
 & 12 \Sigma^3 \p_R d_+ \Sigma
 = - 8\, \Sigma^2 \left(-3 \Sigma^2  + 3\, d_+\Sigma \, \p_R \Sigma \right)
   \\
 &  \qquad + e^{2B}\Big\{ \Sigma^2
   \left[
   4\, d_z \, B \p_RF - 4\, d_z^2 B - 7 \left(d_z B\right)^2 \right.
    \\
   & \qquad\qquad \left. + 2\, \p_Rd_z F + \left( \p_RF\right)^2 \right]
   + 4 \left( d_z \Sigma\right)^2
   \\
 &\qquad\qquad + 2\, \Sigma \left[ d_z\Sigma \left(\p_R F
     - 8\, d_z B \right) - 4\, d_z^2 \Sigma \right]
     \Big\} \,,  \\
     &6 \Sigma^4 \p_R d_+ B
     =  - 9\, \Sigma^3 \left( \p_R \Sigma \, d_+ B
     + \p_R B \, d_+ \Sigma \right)\\
   &\qquad\qquad\quad + e^{2B}\Big\{
   \Sigma^2 \big[ \left(d_zB\right)^2 - d_z B\, \p_R F + d_z^2 B
     \\
 &\qquad\qquad\quad
     - 2\, \p_R d_z F - \left(\p_R F\right)^2 \big]
   -4\, \left( d_z \Sigma \right)^2
  \\
  & \qquad \qquad\quad + \Sigma \left[ d_z\Sigma
    \left( d_z B + 4\, \p_R F \right)
   + 2\, d_z^2 \Sigma \right] 
\Big\}\, , \\
 &6 \Sigma^4 \p_R^2 A = 72 \, \Sigma^2 \, d_+ \Sigma  \, \p_R\Sigma
       - 2 \Sigma^4 \left( 9\, \p_R B \,  d_+ B
       +12 \right)  \\
 & \qquad\quad\quad +3\, e^{2B}
   \left\{
   \Sigma^2
   \left[ 4\, d_z^2 B +7 \left( d_z B \right)^2
     - \left( \p_R F \right)^2  \right]
   \right.  \\
 & \left. \qquad\quad\quad 
   + \,8 \Sigma \left( 2\, d_z B \, d_z \Sigma
   + d_z^2 \Sigma \right) - 4\left(d_z\Sigma\right)^2
\right\}\,,
\end{aligned}
\label{dr_eqs}
\end{equation}
and finally
\begin{align}
\p_\varv B = d_+ B - \tfrac{A}{2} \p_R B\, . \label{dtB_eq}
\end{align}
As in the previous section, there are also two additional equations
that are not explicitly solved. The vector~$d_+$ points to the
direction of the outgoing null rays and hence equations~\eqref{dr_eqs}
do involve derivatives extrinsic to the hypersurfaces of constant
time. However, if one considers~$d_+ B$ and~$d_+ \Sigma$ as
independent variables of the system, then equations~\eqref{dr_eqs} are
intrinsic to the ingoing null hypersurfaces and possess a nested
structure just as in Bondi-gauge. Hence the only equation that
involves derivatives extrinsic to the hypersurfaces of constant
retarded time is~\eqref{dtB_eq}. To analyze the hyperbolicity of the
resulting PDE system we follow exactly the same steps as in the
previous setup.

\subsubsection{First order reduction \& Linearization}

The definition~\eqref{tilde_dot_def} was used earlier to write the
field equations in a more compact form, but for the rest of the
analysis we expand out the definition of~$d_z$. Before performing the
first order reduction, we apply the coordinate transformation~$r=1/R$,
drawing the boundary to~$r=0$. The metric components however still
exhibit singular behavior there, so as elsewhere in the literature,
we apply appropriate field redefinitions to obtain regular fields on
the boundary, namely
\begin{align*}
  A(\varv, r, z)
  &\rightarrow \frac{1}{r^2} + r^2 A(\varv, r, z)
    \,, \\
  B(\varv, r, z)
  &\rightarrow r^4 B(\varv, r, z)
    \,, \\
  \Sigma(\varv, r, z)
  & \rightarrow \frac{1}{r} + r^3 \Sigma (\varv, r, z)
    \,, \\
  F(\varv, r, z)
  &\rightarrow r^2 F(\varv, r, z)
  \,,
\end{align*}
and similarly for derivatives of the above fields. To simplify the
presentation we linearize here about vacuum AdS. Our conclusions are
however unaltered if we work about an arbitrary background. Full
expressions in the general case can be found
in~\cite{GiaHilZil20_public}. We define reduction variables
according to
\begin{align*}
  &A_r = \p_r A \,, B_r = \p_r B \, ,
  F_r = \p_r F \, , \Sigma_r = \p_r \Sigma \,,\\
  &A_z = \p_z A \, , B_z = \p_z B \, ,
  F_z = \p_z F \, , \Sigma_z = \p_z \Sigma \, ,\\
  &B_+ = d_+ B \, , \Sigma_+ = d_+ \Sigma \, .
\end{align*}
The complete first order system, is then
\begin{equation}
\begin{aligned}
    r^4 \p_\varv B &= -S_1 \, , 
    \\
    r^4 \p_\varv B_r&=
    \frac{ r^4}{2} \p_r B_r + r^3 \p_r B_+ - S_2
    \, , 
    \\
    -6 r \p_r B_+
    &= 2 r^2 \p_r F_z + r^2 \p_z B_z + 2 r^2 \p_z \Sigma_z
    - \, S_3 \, ,
    \\
    \p_r B_z
    &= \p_z B_r
    \, , 
    \\
    \p_r \Sigma
    &= -S_5
    \, , 
    \\
    r^7 \p_r \Sigma_r
    &= -S_6 
    \, ,                      
    \\
    12 r  \p_r \Sigma_+
    &=
    2 r^2 \p_r F_z + 4 r^2 \p_z B_z + 8 r^2 \p_z \Sigma_z  - \, S_7
    \, , 
    \\
    \p_r \Sigma_z
    &= \p_z \Sigma_r
    \, ,
    \\
    \p_r F
    &= -S_9
    \, ,
    \\
    r^4 \p_r F_r
    &=
    -4 r^4 \p_r \Sigma_z - 2r^4 \p_r B_z  - \, S_{10} \, , 
    \\
    \p_r F_z
    &= \p_z F_r
    \, ,
    \\
    \p_r A
    &= -S_{12}
    \, ,
    \\
    6 r^2  \p_r A_r
    &=
    12 r^2 \p_z B_z + 24 r^2 \p_z \Sigma_z - \, S_{13} 
    \, ,
    \\
    \p_r A_z
    &= \p_z A_r
    \, ,
  \end{aligned}
  \label{eqn:Lin_Sys_Vac_AdS}
\end{equation}
which can be written as
\begin{align} 
  \mathbfcal{A}^\varv \p_\varv \mathbf{u} + \mathbfcal{A}^r \p_r \mathbf{u}
  + \mathbfcal{A}^z \p_z \mathbf{u}
  + \mathbfcal{S} = 0, \label{AdS_system_matrix_form_1}
\end{align}
with state vector
\begin{align}
  \mathbf{u} = \left( A_r, B_+, \Sigma_+, \Sigma_r, F_r, B_z,
  \Sigma_z, B_r, A_z, F_z, A, F, B, \Sigma \right)^T\,.\nonumber
\end{align}
The principal part matrix associated with the retarded advanced
time~$\mathbfcal{A}^\varv$ is again not invertible and hence we
proceed with a transformation to an appropriate auxiliary frame.

\subsubsection{Coordinate transformation}

To obtain a suitable coordinate frame we transform from~$(\varv,r,z)$
to~$(t,\rho,z)$ with
\begin{align*}
  \varv = t-\rho \, , \qquad r=\rho \, ,
\end{align*}
and the remaining coordinates unaltered, which gives
\begin{align*}
  \p_\varv = \p_t \, ,
  \qquad \p_r = \p_t + \p_\rho  \, ,
\end{align*}
with~$\p_z$ unaffected. Applying the transformation yields
\begin{align*}
    \mathbfcal{A}^t \, \p_t  \mathbf{u} 
    + \mathbfcal{A}^r \, \p_\rho  \mathbf{u}
    + \mathbfcal{A}^z \, \p_z  \mathbf{u} \, + \mathbfcal{S} = 0
    \, ,
\end{align*}
where now $\mathbfcal{A}^t = \mathbfcal{A}^\varv + \mathbfcal{A}^r$ is
invertible. After multiplying from the left with the inverse
of~$\mathbfcal{A}^t$ we again bring the system to the form
\begin{align}
   \p_t  \mathbf{u} 
  + \mathbf{B}^\rho \, \p_\rho  \mathbf{u}
  + \mathbf{B}^z \, \p_z  \mathbf{u} \, + \mathbf{S} = 0
  \, , \label{AdS_new_frame_system_2}
\end{align}
where~$\mathbf{B}^\rho = \left( \mathbfcal{A}^t \right)^{-1}
\mathbfcal{A}^r$ and $\mathbf{B}^z = \left( \mathbfcal{A}^t
\right)^{-1} \mathbfcal{A}^z$. The principal part~$\mathbf{B}^\rho$ is
diagonalizable with real eigenvalues~$0$ and~$\pm 1$. The principal
part~$\mathbf{B}^z$ has the same real eigenvalues but it does not have
a complete set of eigenvectors, so it is not diagonalizable. The
system resulting from this specific first order reduction is thus only
WH. Next, by again constructing generalized characteristic variables
in the~$z$ direction we will examine whether or not an appropriate
addition of the reduction constraints can render the reduction
strongly hyperbolic. The reduction constraints are
\begin{equation}
\begin{aligned}
  &\p_z A - A_z = 0 \,, \qquad 
  \p_z B - B_z = 0 \, ,\\
  &\p_z \Sigma - \Sigma_z = 0 \, ,\qquad 
   \p_z F - F_z = 0 \,,\\
  &\p_zB_\rho - \p_\rho B_z
  = \tfrac{1}{2}\p_zB_r - \p_z B_+ - \p_\rho B_z=0 \,,\\
  &\p_z\Sigma_\rho - \p_\rho \Sigma_z
  = \tfrac{1}{2}\p_z\Sigma_r - \p_z\Sigma_+ - \p_\rho \Sigma_z = 0\,.
\end{aligned}
\label{AN_red_constr_all}
\end{equation}

\subsubsection{Generalized characteristic variables}
\label{Subsubsection:AN_gauge:GCV}

The eigenvalues of~$\mathbf{B}^z$ are~$\lambda = \pm 1$ with algebraic
multiplicity one and~$\lambda =0$ with algebraic multiplicity
twelve. There is one eigenvector for~$\lambda=1$, one for~$\lambda=-1$
and nine for~$\lambda=0$. Since the algebraic and geometric
multiplicity of~$\lambda=0$ differ by three, the Jordan normal
form,
\begin{align*}
  \mathbf{J}^z \equiv \mathbf{T}_z^{-1} \,
  \mathbf{B}^z \, \mathbf{T}_z \,.
\end{align*}
must have some non-trivial block. Let us consider the~$t,z$ part
of~\eqref{AdS_new_frame_system_2} and, as earlier
in~\eqref{gen_char_system}, use~$\mathbf{T}_z^{-1}$ to construct the
generalized characteristic variables in the~$z$ direction,
\begin{align}
\mathbf{v} = \mathbf{T}_z^{-1} \, \mathbf{u}
\end{align}
satisfying
\begin{align}
  &\p_t \mathbf{v} + \mathbf{J}_z  \, \p_z \mathbf{v}
   \simeq 0 \, , \label{tz_gen_char_system}
\end{align}
with~$\simeq$ here denoting equality up to transverse derivatives and
non-principal terms. The components of~$\mathbf{v}$ begin,
\begin{align*}
 & - B_r - \frac{1}{3} B_z - \frac{2}{3}F_r
   -2 \Sigma_r  -\frac{2}{3} \Sigma_z\, ,\\
 & -B_r + \frac{1}{3} B_z + \frac{2}{3} F_r
   - 2 \Sigma_r + \frac{2}{3}\Sigma_z\,,
\end{align*}
with speeds~$\mp 1$ respectively. Next we have those with vanishing
speeds, which are most naturally presented in three blocks. The first
of these consists of the set of {\it true} characteristic variables,
\begin{align*}
  &  B_+ - \frac{\rho}{2} B_r - \rho \Sigma_r\,,\quad
  \Sigma_+ - \frac{\rho}{8} A_r + \frac{\rho}{4} B_r
  + \frac{\rho}{2} \Sigma_r\,,\\
  &\frac{1}{4}A_r + \frac{3}{2} B_r + F_z + 3 \Sigma_r
   \, ,\quad A \, ,\quad  F \, ,\quad B \, ,\quad \Sigma \, ,
\end{align*}
a coupled pair consisting of one generalized and one characteristic
variable, respectively,
\begin{align}
  & -\frac{4}{3}B_z - \frac{2}{3} F_r - \frac{2}{3}\Sigma_z
        \, ,\qquad - 2 \Sigma_r \, ,\label{AAdS_char_var_pair}
\end{align}
and finally a coupled triplet of two generalized characteristic
variables and one characteristic variable, respectively,
\begin{equation}
\begin{aligned}
& \frac{1}{4}A_z + \frac{1}{6} B_z + \frac{1}{3}F_r
  + \frac{1}{3}\Sigma_z \, ,\quad
  -\frac{1}{4}A_r + \frac{1}{2}B_r + \Sigma_r\, , \\
& \frac{2}{3} B_z + \frac{1}{3} F_r
  + \frac{4}{3} \Sigma_z\, .
\end{aligned}
\label{AAdS_char_var_triple}
\end{equation}
In other words, from the structure of the Jordan blocks
of~$\mathbf{J}^z$, reading off the components
of~\eqref{tz_gen_char_system} the first member of the
pair~\eqref{AAdS_char_var_pair} and the first two members of the
triple~\eqref{AAdS_char_var_triple} we have the schematic form,
\begin{align}
  \p_t v_i + \p_z v_{i+1} \simeq 0 \,, \label{Jordan_pathology}
\end{align}
with~$v_i$ referring to the field and~$v_{i+1}$ the next element of
the pair or triple. The question is whether or not there exists an
appropriate addition of the reduction
constraints~\eqref{AN_red_constr_all} such that equations of the
form~\eqref{Jordan_pathology} are turned into equations of the form
\begin{align}
  \p_t v_i + \lambda_i \, \p_z v_i \simeq 0 \,,
  \label{jordan_pathology_cured}
\end{align}
where we are allowing different first order reductions to adjust also
characteristic speeds. This is a necessary condition for building an
alternative reduction that is SH. This would mean that the generalized
characteristic variable~$v_i$ that is originally coupled
with~$v_{i+1}$ could be decoupled, and the respective generalized
eigenvector replaced by a simple eigenvector. We examine this for the
second two elements of the triplet~\eqref{AAdS_char_var_triple} and
show by contradiction that this necessary condition can not be
fulfilled. With our original, specific reduction we have
\begin{equation}
\begin{aligned}
  & \p_t \left(
    \frac{2}{3} B_z + \frac{1}{3} F_r + \frac{4}{3} \Sigma_z\right)
    \simeq 0 \, , \\
  & \p_t \left( -\frac{1}{4}A_r + \frac{1}{2}B_r + \Sigma_r \right)
    \\
  & \qquad \qquad \qquad +
    \p_z \left(
    \frac{2}{3} B_z + \frac{1}{3} F_r + \frac{4}{3} \Sigma_z\right)
    \simeq 0 \,.
  \end{aligned}
  \label{tz_gen_char_eq12}
\end{equation}
Observe, first of all, that neither of these two equations, nor the
two large terms grouped separately in the second, can be written as a
linear combination (equality taken here in the sense of~$\simeq$) of
the reduction constraints~\eqref{AN_red_constr_all}. The choice of
reduction lies in the freedom to add multiples of the six reduction
constraints~\eqref{AN_red_constr_all} to the evolution
equations. Suppose that some choice of addition of these constraints
did result in a SH first order reduction. Starting with the first
equation of~\eqref{tz_gen_char_eq12}, for our alternative reduction we
have
\begin{align}
 \p_t \left(
 \frac{2}{3} B_z + \frac{1}{3} F_r
 + \frac{4}{3} \Sigma_z\right)
 \simeq \sum_\alpha c_\alpha\,C_\alpha\,,\label{eqn:alternative_reduction_cv}
\end{align}
with the terms on the right-hand-side a linear combination of the
reduction constraints~$C_\alpha$. Since this alternative reduction is
SH we have,
\begin{align*}
  \sum_\alpha c_\alpha\,C_\alpha\simeq \sum_\alpha a^{0}_\alpha \p_zv^0_\alpha
  +\sum_\alpha a^{\pm}_\alpha \p_zv^{\pm}_\alpha\,,
\end{align*}
with~$v^0_\alpha$ denoting the set of~$0$-speed characteristic
variables and~$v^{\pm}_\alpha$ denoting the remaining characteristic
variables.
Using~$\p_tv^{\pm}_\alpha\simeq\lambda_\alpha \p_zv^{\pm}_\alpha$ we
may therefore rewrite~\eqref{eqn:alternative_reduction_cv} as
\begin{align*}
 \p_t \left( \frac{2}{3} B_z + \frac{1}{3} F_r + \frac{4}{3}
 \Sigma_z-\sum_\alpha a^{\pm}_\alpha\lambda_\alpha^{-1}
 v^{\pm}_\alpha\right) \simeq
 \sum_\alpha a^{0}_\alpha \p_zv^0_\alpha.
\end{align*}
Now, by our observation directly after~\eqref{tz_gen_char_eq12}, the
term inside the large bracket can not vanish identically. Therefore we
must have~$a^{0}_\alpha=0$ or we have found, on the left-hand-side, a
non-trivial generalized characteristic variable, in contradiction to
the assumption that our reduction is SH. Moving on to the second
equation of~\eqref{tz_gen_char_eq12}, we can write the equivalent
expression for the alternative first order reduction as,
\begin{align*}
  & \p_t \left( -\frac{1}{4}A_r + \frac{1}{2}B_r + \Sigma_r \right) +
  \p_z \left( \frac{2}{3} B_z + \frac{1}{3} F_r + \frac{4}{3}
  \Sigma_z\right) \\ &\simeq\sum_\alpha
  c'_\alpha\,C_\alpha\,,
\end{align*}
again with the right-hand-side a linear combination of the reduction
constraints. From here a simple calculation shows that
\begin{align*}
  -\frac{1}{4}A_r + \frac{1}{2}B_r + \Sigma_r
  +\sum_\alpha a'_\alpha \lambda_\alpha^{-1}v^{\pm}_\alpha\,,
\end{align*}
is nevertheless {\it still} a non-trivial generalized characteristic
variable for a suitable choice of~$a'_\alpha$. By contradiction we
have therefore shown that there is no first order reduction that gives
a SH first order PDE system in the~$(t,\rho,z)$ frame used here.

\subsection{Frame independence}\label{Subsection:frame_independence}

In the previous subsections we presented a hyperbolicity analysis of
two widely used Bondi-like formulations of GR. We worked with a
particular auxiliary frame with one timelike element and the remainder
spacelike. The auxiliary basis was used to express the original PDEs,
which were then shown to be only WH. In this subsection we argue that
this result persists for other auxiliary frames. Our argument is based
on the dual foliation (DF) approach of~\cite{Hil15} and follows
closely Sec.~II.D of~\cite{SchHilBug17}. In this subsection, Latin
letters~$a\dots e$ are used as abstract indices, Greek letters run
from~$0$ to~$d+1$ for a~$d+1$-dimensional spacetime and a given basis
and Latin indices~$i,j,k$ denote only the spatial components of this
basis. We also use~$p$ as an abstract index for the spatial
derivatives appearing on the right hand side of a first order PDE. The
symbol~$\p_\alpha$ stands for the flat covariant derivative naturally
defined by~$x^\mu$.

The idea of the DF approach is to express a region of spacetime in
terms of two different frames, which we call uppercase and lowercase.
Considering a~$d+1$ split of the spacetime, let us denote as~$n^a$
and~$N^a$ the normal vectors on the hypersurfaces of constant time for
the lower and uppercase frames, respectively. We call~$v^a$ and~$V^a$
the boost vectors for each frame, which are spatial with respect to
the corresponding normal vector. The Lorentz factor is~$W=\left(1- v^a
v_a\right)^{-1/2}=\left(1- V^a V_a\right)^{-1/2}$ and we denote
as~$\gamma_{ab}$ and~$^{(N)}\!\gamma_{ab}$ the lower and uppercase
spatial metrics. The following useful relations hold
\begin{equation}
  \begin{aligned}
  \delta^a{}_b &= \gamma^a{}_b-n^a n_b
  = {}^{(N)}\!\!\gamma^a{}_b - N^aN_b \,,\\
  n^a & = W \left(N^a+V^a\right)\,,
  \quad N^a = W \left(n^a+v^a\right) \,.
\end{aligned}
  \label{low_up_projectors}
\end{equation}
Let us consider a first order PDE in the compact form
\begin{align*}
  \mathbfcal{A}^b \delta^a{}_b \partial_a \mathbf{u}
  + \mathbfcal{S} = 0 \, ,
\end{align*}
and~$d+1$ split using the lower and uppercase frames,
replacing~$\delta^a{}_b$ by means of~\eqref{low_up_projectors},
giving
\begin{align}
  \mathbfcal{A}^n  \partial_n \mathbf{u}
  &\simeq \mathbfcal{A}^b \gamma^a{}_b \partial_a \mathbf{u}
    \,, \quad
  \mathbfcal{A}^N  \partial_N \mathbf{u}
  \simeq \mathbfcal{A}^b \,^{(N)}\!\gamma^a{}_b \partial_a \mathbf{u}
    \, . \label{lowercase_uppercase_PDE}
\end{align}
We obtain two evolution systems for the variables of~$\mathbf{u}$,
with
\begin{equation}
\begin{aligned}
  \mathbfcal{A}^a n_a
  &\equiv \mathbfcal{A}^n \, ,
  \quad
  n^a \partial_a \equiv \partial_n
  \, ,\\
  \mathbfcal{A}^a N_a
  &\equiv \mathbfcal{A}^N \, ,
  \quad
  N^a \partial_a \equiv \partial_N\, .
\end{aligned}
  \label{up_low_frames_matrix_defs}
\end{equation}
Without loss of generality we choose to identify the uppercase frame
with the auxiliary frames used in
subsections~\ref{Subsection:BS_gauge}
and~\ref{Subsection:AN_gauge}. The definitions
\begin{align*}
  \mathbfcal{A}^n &\equiv \mathbf{A}^n \, ,
  \quad \quad \; \,
  \mathbfcal{A}^a \, \gamma^b{}_a \equiv
  \mathbf{A}^b \, , \\
  \mathbfcal{A}^N & \equiv \mathbf{B}^N \, ,
  \quad
  \mathbfcal{A}^a \, ^{(N)}\!\gamma^b{}_a \equiv
  \mathbf{B}^b \, , 
\end{align*}
imply~$\mathbf{B}^b N_b=0$,~$\mathbf{A}^b n_b=0$ and lead to the
following upper and lowercase first order PDE forms
\begin{align}
  \p_N \mathbf{u} &= \mathbf{B}^p \p_p \mathbf{u} - \mathbf{S}
  \,, \quad
  \mathbf{A}^n \p_n \mathbf{u} = \mathbf{A}^p \p_p \mathbf{u} - \mathbf{S}
  \,, \label{uppercase_lowercase_new_frame_PDE}
\end{align}
where~$\mathbf{B}^N = \mathbb{1}$ by assumption. The former is the
same form as in equations~\eqref{flat_new_frame_system_2}
and~\eqref{AdS_new_frame_system_2}. In this form we found the PDE
systems only WH due the~$2\times2$ Jordan blocks of the angular
principal parts. This can be represented in a generalized eigenvalue
problem of the form
\begin{align}
  \mathbf{l}^N_{\lambda_N}
  \left(\mathbf{P}^S - \mathbb{1} \lambda_N\right)^M
  = 0 \, , \label{up_gen_eigenv_prob}
\end{align}
where~$S^a$ is a unit spatial vector,~$\mathbf{P}^S \equiv
\mathbf{B}^a S_a$ the principal symbol and~$M$ is the rank of the
generalized left eigenvector~$\mathbf{l}^N_{\lambda_N}$ with
eigenvalue~$\lambda_N$, with~$M=2$ for the generalized eigenvectors
that correspond to the aforementioned Jordan blocks. We wish to
examine if generalized eigenvalue problems of this form exist also in
the lowercase frame. Hence we need to relate the two equations
of~\eqref{uppercase_lowercase_new_frame_PDE}, obtaining
\begin{equation}
\begin{aligned}
  \mathbf{A}^n&=W(\mathbb{1} + \mathbf{B}^V)\,,\\
  \mathbf{A}^p&= \mathbf{B}^a(\gamma^p{}_a+WV_av^p)
                -W(\mathbb{1} + \mathbf{B}^V)v^p\,,
\end{aligned}
\label{low_in_up_PDE}
\end{equation}
and
\begin{equation}
\begin{aligned}
  \mathbf{B}^N&=\mathbb{1}=W(\mathbf{A}^n + \mathbf{A}^v)\,,\\
  \mathbf{B}^p&= \mathbf{A}^a \, ^{(N)}\!\gamma^p{}_a -
                W \mathbf{A}^n V^p \,,
\end{aligned}
\label{up_in_low_PDE}
\end{equation}
where we write~$\mathbf{B}^aV_a\equiv\mathbf{B}^V$. Let us
examine~$\mathbb{1} + \mathbf{B}^V$. In~\cite{SchHilBug17}
invertibility of this matrix was guaranteed by strong
hyperbolicity. Here we want to analyze PDEs that are only WH and so
may not assume that~$\mathbf{B}^V$ is diagonalizable. Hence, let us
denote as
\begin{align*}
\mathbf{J}^{S_V} = \mathbf{T}^{-1}_{S_V} \mathbf{B}^{S_V} \mathbf{T}_{S_V}\,,
\end{align*}
the Jordan normal form of~$\mathbf{B}^{S_V}=\mathbf{B}^a(S_V)_a$,
where~$V^a=|V|S_V^a$ is the uppercase boost vector with norm~$|V|$
pointing in the direction of~$S_V^a$. One can write each
block~$\mathbf{j}$ of the Jordan form~$\mathbf{J}$ with only the
eigenvalue~$\lambda_i$ on the diagonal as
\begin{align*}
\mathbf{j} = \lambda_i \mathbb{1} + \mathbf{N}\,,
\end{align*}
where~$\mathbf{N}$ is a nilpotent matrix of the size of~$\mathbf{j}$
with~$\mathbf{N}^q=0$. Consequently
\begin{align*}
  \mathbf{T}^{-1}_{S_V}\left(\mathbb{1} + \mathbf{B}^V\right)
  \mathbf{T}_{S_V}=\mathbb{1} + \mathbf{J}^{S_V}|V|\,,
\end{align*}
and for each block~$\mathbf{j}^{S_V}$,
\begin{align*}
  \mathbb{1} + \mathbf{j}^{S_V}
  =\tilde{\lambda}^{S_V}_i \left(\mathbb{1}
  + \frac{|V|}{\tilde{\lambda}^{S_V}_i}\mathbf{N}^{S_V}\right)\,,
\end{align*}
assuming that
\begin{align}
  \tilde{\lambda}^{S_V}_i = 1 + |V|\lambda_i^{S_V}
  \neq 0 \,. \label{lambda_tilde_condition}
\end{align}
The inverse of this block is then
\begin{align*}
  \frac{1}{\tilde{\lambda}^{S_V}_i}
  \left[
  \mathbb{1} + \sum_{j=1}^{q-1}
  \left(-\frac{|V|}{\tilde{\lambda}^{S_V}_i}\right)^j
  \left(\mathbf{N}^{S_V}\right)^j
  \right]\,,
\end{align*}
and hence~$\mathbb{1}+\mathbf{B}^V$ is invertible as long as
condition~\eqref{lambda_tilde_condition} is satisfied for
each~$\lambda_i$. Note that in our normalization light-speed
corresponds to~$\lambda=1$. Since~$|V|<1$,
inequality~\eqref{lambda_tilde_condition} is always satisfied for
physical propagation speeds, although could be violated when
superluminal gauge speeds are present. If one considers for instance
the analysis of subsections~\ref{Subsection:BS_gauge}
and~\ref{Subsection:AN_gauge} on top of Minkowski and vacuum AdS
background respectively, then this condition is satisfied. We wish to
find the equivalent of the uppercase generalized eigenvalue
problem~\eqref{up_gen_eigenv_prob} in the lowercase frame. Thus, using
the second equation of~\eqref{up_in_low_PDE} and~$S_a=\s_a-W
V^Sn_a$~\cite{SchHilBug17,HilSch18} we express the principal symbol in
the lowercase frame, namely
\begin{align*}
\mathbf{P}^S& \equiv \mathbf{B}^a S_a
=\mathbf{A}^a \s_a - \mathbf{A}^nWV^S \,.
\end{align*}
Hence, the equivalent of~\eqref{up_gen_eigenv_prob} in the lowercase
frame is
\begin{align}
  \mathbf{l}^N_{\lambda_N}
  \left[ \mathbf{A}^{(\s-\lambda_N Wv)}
    -W(\lambda_N+V^S)\mathbf{A}^n\right]^M = 0 \,.
  \label{low_gen_eigenv_prob}
\end{align}
Thus if in the uppercase frame the
eigenproblem~\eqref{up_gen_eigenv_prob} with~$M=1$ fails to admit a
complete set of left eigenvectors then so does the lowercase frame,
and so both setups would be at best weakly hyperbolic. To see this we
need only set~$M=1$ in~\eqref{low_gen_eigenv_prob} and note that the
lowercase principal symbol in the~$\s_a-\lambda_N Wv_a$ direction is
proportional to
\begin{align*}
  (\mathbf{A}^n)^{-1}\mathbf{A}^{(\s-\lambda_N Wv)}\,.
\end{align*}
and so deficiency of the lower case principal symbol in this direction
is equivalent to that of the upper case principal symbol stated
before. Unfortunately the relationship between the upper and lowercase
generalized left eigenvectors is more subtle.  Returning to our
specific systems and identifying the uppercase unit spatial
vector~$S^a$ with the unit spatial vectors in the~$\partial_\theta$
and~$\partial_z$ directions of
subsection~\ref{Section:characteristic_formulations}, we conclude that
weak hyperbolicity of those PDEs persists in other frames.

\section{Toy models} \label{Section:toy_models}

In this section we introduce two toy models, one SH and one WH, which
capture the core structure of the systems analyzed in the previous
section. Our aim is to examine the consequence of the algebraic
properties determined earlier on local well-posedness in the context
of the CIBVP. The principal parts of the two models differ only in the
angular direction~$z$, with the WH model possessing a
non-diagonalizable principal symbol.

\subsection{The PDEs} \label{Subsection:the_models}

The equations of motion for the WH model are,
\begin{equation}
\begin{aligned}
  & \p_x \phi = -S_\phi
    \, ,\\
  & \p_x \psi_\varv  - \p_z \phi = -S_{\psi_\varv} 
    \, ,\\
  & \p_u \psi - \frac{\left(1-x^2\right)^{3/2}}{2 c_x}  \p_x \psi
    -\p_z\psi = -S_\psi \,,
\end{aligned}
\label{eqn:WH_model}
\end{equation}
with~$x \in [0,1]$,~$z \in [0,2\pi)$ with periodic boundary
conditions, $u \geq u_0$ for some initial time~$u_0$ and~$c_x$ a
constant. This PDE can be written in the form
\begin{align}
  \mathbf{A}^u \, \p_u \mathbf{u} + \mathbf{A}^x \, \p_x \mathbf{u}
  + \mathbf{A}^z \, \p_z \mathbf{u} + \mathbf{S} = 0\, ,
\end{align}
where~$\mathbf{u} = (\phi, \psi_\varv, \, \psi )^T$ is the state vector,
and the principal matrices are given by
\begin{align*}
  \mathbf{A}^u &= \mbox{diag}(0,0,1)\\
  \mathbf{A}^x &= \mbox{diag}(1,1,\tfrac{-1}{2 c_x}(1-x^2)^{3/2})
\end{align*}
and
\begin{align*}
  \mathbf{A}^z = \begin{pmatrix}
    0 & 0 & 0\\
   -1 & 0 & 0\\
    0 & 0 & -1 
  \end{pmatrix}\,.
\end{align*}
The source terms are denoted by~$S_\phi, S_{\psi_\varv}$
and~$S_\psi$. The first two Eqs.\ of~\eqref{eqn:WH_model} are intrinsic
to a hypersurface of constant~$u$, whereas the last is the ``evolution
equation'' of the system. The angular principal part~$\mathbf{A}^z$ is
not diagonalizable since it has a~$2\times2$ Jordan block for the
intrinsic equations, mimicking the core structure of the previously
analyzed single-null PDEs. One may think of this model as a simplified
analog of these systems with a compactified radial coordinate, similar
to the way that the Bondi-Sachs formulation is used for characteristic
extraction. This role can be played by the coordinate~$x$ with~$c_x$ a
constant involved in the compactification. More specifically
\begin{align*}
  x = \frac{r-r_\textrm{min}}{\sqrt{c_x^2 + (r-r_\textrm{min})^2}} \,,
\end{align*}
where~$r_\textrm{min}$ is the minimum physical radius that we consider
and the factor~$c_x$ controls the density of points towards~$r
\rightarrow \infty$, if we were to map the compactified grid~$x$ to
the physical radius grid~$r$.

By removing the angular derivative from the second intrinsic
equation~\eqref{eqn:WH_model} we obtain our SH toy model
\begin{equation}
\begin{aligned}
  & \p_x \phi = -S_\phi
        \, ,\\
  & \p_x \psi_\varv  = -S_{\psi_\varv} 
        \, ,\\
  & \p_u \psi - \frac{\left(1-x^2\right)^{3/2}}{2 c_x}  \p_x \psi
    -\p_z\psi = -S_\psi
    \, ,
  \end{aligned}
  \label{eqn:SH_model}
\end{equation}
which has the same principal part matrices~$\mathbf{A}^u$
and~$\mathbf{A}^x$ as before, but has diagonal~$\mathbf{A}^z$. We
employ this model for comparison between numerical results with SH and
WH systems. The PDE problem for both systems~\eqref{eqn:WH_model}
and~\eqref{eqn:SH_model} has as domain
\begin{align*}
  x \in [0,1]
  \,, \quad
  z \in [0,2\pi)
  \,, \quad
  u \in [u_0, u_f]
  \,,
\end{align*}
for some initial and final times~$u_0$ and~$u_f$ respectively. We
apply periodic boundary conditions in the~$z$ direction for
simplicity. The initial and boundary data are
\begin{align}
  \psi_* \equiv \psi(u_0,x,z)
\end{align}
and
\begin{align}
  \hat{\phi} \equiv \phi(u,0,z)
  \,, \qquad
  \hat{\psi_\varv} \equiv \psi_\varv(u,0,z)
  \,,\label{eqn:model_boundary_data}
\end{align}
respectively and are freely specifiable.

\subsection{Algebraic determination of
  well-posedness} \label{Subsection:algebraic_characterization}

So far we have discussed the degree of hyperbolicity of GR in two
gauges and constructed models that capture the basic structure we
unearthed. As mentioned in the Introduction the reason we care about
this algebraic characterization is that, in the linear constant
coefficient approximation, it determines well-posedness of the initial
value problem~\cite{KreLor89,GusKreOli95}. In this subsection we
present our well-posedness analysis, focusing on the WH toy
model. The interested reader can find the complete analysis of both
our models in~\cite{GiaHilZil20_public}. In this analysis we work in
the constant-coefficient approximation, following closely the
philosophy and notation of~\cite{KreLor89}. We start with the IVP and
adjust our results to the CIBVP at the end. Specifically, we wish to
understand what inequalities, with what norms, can be used to bound
solutions in terms of their given data, and how lower order
perturbations affect such estimates.

Consider the Cauchy problem for the linear, constant coefficient
system,
\begin{align}
  \p_t\mathbf{u}&=\mathbf{B}^p\p_p\mathbf{u}+\mathbf{S}
  \equiv\mathbf{B}^p\p_p\mathbf{u}+\mathbf{B}\mathbf{u}\,.
\end{align}
To be well-posed in the~$L^2$-norm we must have real
constants~$K\geq1$ and~$\alpha \in \mathbb{R}$ such that
\begin{align}
  | e^{\mathbf{P}(i \omega) t}|
  \leq K e^{\alpha t} \, , \label{wp_inequality}
\end{align}
for all~$t\geq0$ and all~$\omega \in \mathbb{R}^n$. Here
\begin{align}
  \mathbf{P}(i \omega) =  i \omega_p \mathbf{B}^p + \mathbf{B}
\end{align}
is the constant-coefficient symbol of the PDE after Fourier
transforming in space, with~$i\omega_p\mathbf{B}^p$ the principal
symbol and~$\mathbf{B} \, \mathbf{u} = -\mathbf{S}$ the lower order
term related to sources. Essentially, inequality~\eqref{wp_inequality}
states that the solution of the PDE has to be bounded at each time by
an exponential that is independent of the Fourier mode~$\omega_p$. In
this manner we obtain an estimate of the solution~$\mathbf{u}$ at all
times by the initial data~$f$
\begin{align*}
  || \mathbf{u}( \cdot \, ,t) ||_{L^2}
  &= || e^{\mathbf{P} (i \omega) t } \hat{f}(\omega) ||_{L^2}\\
  & \leq
  K e^{\alpha t} || \hat{f} ||_{L^2} = K e^{\alpha t} ||f||_{L^2}\, .
\end{align*}
In the terminology of~\cite{KreLor89}, if a Cauchy problem instead
satisfies only
\begin{align}
  | e^{\mathbf{P}(i \omega) t}|
  \leq K_1 e^{\alpha t}  \left(1 + | \omega|^q  \right)\, ,
  \label{weak_wp_inequality}
\end{align}
with~$q$ some natural number, it is called weakly well-posed. This
type of estimate is weaker than~\eqref{wp_inequality}, because the
explicit appearance of~$\omega$ on the right hand side makes it
impossible to bound the solution by an exponential independent
of~$\omega$. If, rather than insisting on~$L^2$ we allow also some
{\it specific} derivative, determined by the system, within the norm,
we can nevertheless obtain the estimate
\begin{align*}
  || \mathbf{u}( \cdot \, ,t) ||_q \leq K_2 \, e ^{\alpha t} ||f||_q \, .
\end{align*}
for the solution~$\mathbf{u}$. This would not be terrible, except that
if the PDE is only weakly well-posed, then perturbations to the system
by generic lower order terms will lead to frequency dependent
exponential growth of the solution, and the resulting perturbed
problem is ill-posed in any sense. We show this explicitly for our WH
models later. The latter is not true for well-posed problems, which
remain well-posed in the presence of lower order
perturbations~\cite{SarTig12, KreLor89}.

To apply the above results directly the system needs to be written in
a form where the time principal part is the identity matrix. We
achieve the latter via a coordinate transformation similar to those of
Sec.~\ref{Section:characteristic_formulations},
\begin{align*}
  u=t-\rho \, , \qquad  x = \rho \, ,  \qquad  z = z\,.
\end{align*}
Starting from Eqs.~\eqref{eqn:WH_model}, we bring the system to
the form
\begin{align*}
  \p_t \phi & = - \p_\rho \phi  - S_\phi
    \, ,\\
  \p_t \psi_\varv  &= - \p_\rho \psi_\varv + \p_z \phi -  S_{\psi_\varv} 
    \, ,\\
  \p_t \psi &= F \, \p_\rho \psi
    + G  \, \p_z\psi - G \, S_\psi\,,
\end{align*}
where
\begin{align*}
  F = \frac{\left( 1 - \rho^2 \right)^{3/2}}
  {2 c_x - \left( 1 - \rho^2 \right)^{3/2}}\,,
  \quad
  G = \frac{2 c_x}{2 c_x - \left( 1 - \rho^2 \right)^{3/2}}
\end{align*}
are fixed real constants for fixed~$\rho$ and~$c_x$, with non-zero
denominator for our~$\rho$ domain and an appropriately
chosen~$c_x$. In this frame the principal parts
are~$ \mathbf{B}^t=\mathbf{1}$ and
\begin{align*}
  \mathbf{B}^\rho = \begin{pmatrix}
    -1 & 0 & 0\\
    0 & -1 & 0\\
    0 & 0 & F 
  \end{pmatrix}
  ,\qquad
  \mathbf{B}^z = \begin{pmatrix}
    0 & 0 & 0\\
    1 & 0 & 0\\
    0 & 0 & G 
  \end{pmatrix}
  .
\end{align*}
This is the auxiliary Cauchy-type setup for the WH model, similarly to
the PDEs in section~\ref{Section:characteristic_formulations} after
the coordinate transformation. After applying a Fourier
transformation, the principal symbol for the WH model is
\begin{align*}
  i\omega_p\mathbf{B}^p = i \omega_\rho \mathbf{A}^\rho + i \omega_z
  \mathbf{A}^z \, .
\end{align*}

\subsubsection{Homogeneous WH model} \label{Subsubsection:homogeneous_WH}

Focusing first on the homogeneous WH model where~$S_\phi =
S_{\psi_\varv} = S_\psi =0$, we obtain
\begin{align}
& e^{(i\hat{\omega}_p\mathbf{B}^p)|\omega| t}=\nonumber\\
&  \begin{pmatrix}
    e^{-i |\omega| \hat{\omega}_\rho t}                   & 0               & 0\\
    i |\omega| \hat{\omega}_z t \, e^{-i |\omega| \hat{\omega}_\rho t}  & e^{-i |\omega| \hat{\omega}_\rho t} & 0\\
    0                                 & 0               & e^{i |\omega| (F \hat{\omega}_\rho + G \hat{\omega}_z) t} 
  \end{pmatrix}
  \, , \label{exp_matrix_homogeneous_WH}
\end{align}
where we express the wavevector as
\begin{align*}
  \omega_p = |\omega| \hat{\omega}_p\,,
\end{align*}
with~$|\omega|$ its magnitude so that~$\hat{\omega}_\rho^2 +
\hat{\omega}_z^2 =1$. The norm of~\eqref{exp_matrix_homogeneous_WH} is
(see chapter~2 of~\cite{SarTig12} for useful definitions)
\begin{align}
  \left|e^{(i\hat{\omega}_p\mathbf{B}^p)|\omega| t}\right|^2 & = 1
  + \frac{|\omega|^2 \hat{\omega}_z^2 t^2}{2} \nonumber \\
& \quad
  + \left[ \left(1 + \frac{|\omega|^2 \hat{\omega}_z^2 t^2}{2} \right)^2
  - 1  \right]^{1/2}. \label{homogeneous_WH_norm}
\end{align}
This norm behaves as~$|\omega| t$ for large~$|\omega|$ and so the
homogeneous WH model obeys an inequality of the
form~\eqref{weak_wp_inequality}, with~$q=1$. Hence, this PDE is only
weakly well-posed, and so satisfies an estimate in
some~$||\cdot||_q$-norm. This norm is specified for our system in
Sec.~\ref{Subsubsection:CCMvsCCE}. If one would discard from the
previous analysis the equation for~$\psi_\varv$ of the homogeneous WH
model~\eqref{eqn:WH_model} since it is decoupled, the remaining
subsystem would be symmetric hyperbolic and one might expect
well-posedness of the full system in the~$L^2$-norm. However, as shown
in Fig.~\ref{fig:L2_norms_all}, this expectation is not true.

\subsubsection{Inhomogeneous WH model} \label{Subsubsection:inhomogeneous_WH}

For the homogeneous WH model we computed the norm
of~$e^{(i\hat{\omega}_p\mathbf{B}^p)|\omega| t}$ to estimate the
behavior of solutions. However, we could also examine the form of the
eigenvalues of the full symbol~$\mathbf{P}(i \omega)$ for
large~$|\omega|$ to understand if the solutions exhibit exponential
growth in~$\omega_p$ (see lemma 2.3.1 of~\cite{KreLor89}). If there is
any eigenvalue~$\lambda$ of~$\mathbf{P}(i \omega)$ such that
\begin{align*}
  \Re[\lambda] \sim |\omega|^s > 0 \; \text{with} \; s>0 \, ,
\end{align*}
for large~$|\omega|$, then solutions of the PDE may exhibit frequency
dependent exponential growth, and the PDE problem is ill-posed in any
sense. For the inhomogeneous WH model we consider the following
possible lower order source terms
\begin{align*}
  \mathbf{B}_1 =
  \begin{pmatrix}
    0 & 0 & 1\\
    1 & 0 & 1\\
    1 & 0 & 0 
  \end{pmatrix}
  ,\,
  \mathbf{B}_2 =
  \begin{pmatrix}
    1 & 0 & 1\\
    1 & 1 & 1\\
    1 & 1 & 1 
  \end{pmatrix}
  ,\,
  \mathbf{B}_3 =
  \begin{pmatrix}
    0 & 1 & 0\\
    0 & 0 & 0\\
    0 & 0 & 0 
  \end{pmatrix}
  \, ,
\end{align*}
where~$-\mathbf{S} = \mathbf{B}\mathbf{u}$. The choice~$\mathbf{B}_1$
is motivated by analogy with the linearized Bondi-Sachs system
with~$\phi \sim \beta$,~$\psi_\varv \sim V$ and~$\psi \sim
\gamma_r$. In~$\mathbf{B}_2$ we include all possible source terms that
do not break the nested structure of the intrinsic equations and
finally in choice~$\mathbf{B}_3$ we introduce source terms that
violate the nested structure, thus rendering the intrinsic system a
coupled PDE. For both~$\mathbf{B}_1$ and~$\mathbf{B}_2$ the
eigenvalues of~$\mathbf{P}(i \omega)$ are
\begin{align*}
  \lambda_1 = \lambda_2 = - i |\omega| \, \hat{\omega}_\rho\,,
  \quad
  \lambda_3 = i |\omega|
  \left( F\, \hat{\omega}_\rho + G \,\hat{\omega}_z \right) \, ,
\end{align*}
as~$|\omega| \rightarrow \infty$, with the next terms appearing at
order~$|\omega|^0$. For these choices of lower order source terms the
inhomogeneous WH model remains well-posed in the lopsided norm. On the
other hand if~$\mathbf{B} = \mathbf{B}_3$ the eigenvalues of the
symbol are
\begin{align*}
  \lambda_1 & =  - i |\omega| \hat{\omega}_\rho
  - (-1)^{1/4} \sqrt{|\omega| \hat{\omega}_z} + O(|\omega|^0) \\
  \lambda_2 & = - i |\omega| \hat{\omega}_\rho
  + (-1)^{1/4} \sqrt{|\omega| \hat{\omega}_z} + O(|\omega|^0) \\
  \lambda_3 & = i |\omega| \left( F\, \hat{\omega}_\rho
  + G\,\hat{\omega}_z \right)
  + O(|\omega|^0)\, .
\end{align*}
for large~$|\omega|$. Since~$\Re[\lambda] \sim |\omega|^{1/2}$, we
conclude that when the nested structure of the intrinsic equations is
broken, the solution of the inhomogeneous WH exhibits frequency
dependent exponential growth. Consequently, the IVP with this system
is no longer weakly well-posed but ill-posed. Note, in contrast, that
for the homogeneous SH model we have
\begin{align*}
  | e^{\mathbf{P}(i \omega) t} | = 1.
\end{align*}
Hence for this model, the IVP is well-posed already in the~$L^2$
norm. Unlike the WH model, well-posedness for this model is not
affected by source terms.

\subsubsection{The CIBVP, CCE and CCM} \label{Subsubsection:CCMvsCCE}

\begin{figure}[t]
  \includegraphics[width=0.33\textwidth]{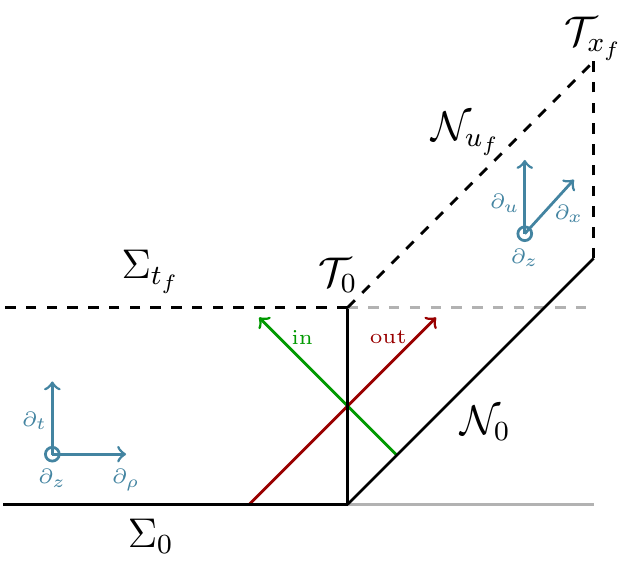}
  \caption{The IBVP (left) and the CIBVP (right) setups.  For CCE
    outgoing data from the IBVP serve as boundary data
    on~$\mathcal{T}_0$ for the CIBVP, which can be viewed as an
    independent PDE problem. In this case the IBVP's spatial domain is
    more extended such that data on $\mathcal{T}_0$ are unaffected by the
    boundary conditions chosen for the problem.  For CCM the IBVP and
    CIBVP are solved simultaneously and out/ingoing data are
    communicated from one to the other
    via~$\mathcal{T}_0$. Effectively, the two problems are viewed as
    one.
    \label{FIG:CCM}
  }
\end{figure}

The previous analysis was performed in Fourier space and yielded that
an IVP based on the homogeneous WH model may be well-posed in an
appropriate lopsided norm, whereas one on the SH model is (strongly)
well-posed in the~$L^2$-norm. We now present our energy estimates for
solutions to the IBVP and CIBVP by working in position space. For
concreteness and simplicity the PDE system for the IBVP is a
homogeneous SH model (which is furthermore symmetric hyperbolic)
\begin{align}
  \begin{aligned}
    & \p_t \bar{\phi} + \p_\rho \bar{\phi} + \p_z\bar{\psi_\varv} = 0 \,,
    \\
    & \p_t \bar{\psi_\varv} + \p_\rho \bar{\psi_\varv}+\p_z\bar{\phi}  = 0 \,,
    \\
    & \p_t \bar{\psi} - \tfrac{1}{2} \p_\rho \bar{\psi} - \p_z \bar{\psi}  = 0 \,,
  \end{aligned}\label{eqn:CCM_IBVP_sys}
\end{align}
with initial data~$\bar{\phi}_*,\, \bar{\psi_\varv}_*,\, \bar{\psi}_*$
on~$\Sigma_0$, boundary data~$\hat{\bar{\psi}}$ on~$\mathcal{T}_0$ and
domain~$t \in[0,t_f],\, \rho \in (-\infty,0]$ and the
compact~$z \in [0, 2\pi)$, and for the CIBVP the homogeneous WH model
\begin{subequations}
  \begin{align}
    &  \p_x \phi  = 0 \,,   \label{eqn:CCM_CIBVP_sys_a}
    \\
    & \p_x \psi_\varv - \p_z\phi  = 0 \,,   \label{eqn:CCM_CIBVP_sys_b}
    \\
    & \p_u \psi - \tfrac{1}{2} \p_x \psi - \p_z \psi  = 0 \,,
    \label{eqn:CCM_CIBVP_sys_c}
  \end{align}
  \label{eqn:CCM_CIBVP_sys}
\end{subequations}
with initial data~$\psi_*$ on~$\mathcal{N}_0$, boundary
data~$\hat{\phi}$ and~$\hat{\psi_\varv}$ on~$\mathcal{T}_0$ and
domain~$u \in[0,u_f],\, x \in [0,x_f]$ and the aforementioned~$z$. The
domains of the two problems are illustrated in Fig.~\ref{FIG:CCM}. We
view the IBVP as a simplified analog of GR in strongly (here even
symmetric) hyperbolic formulations widely used in Cauchy-type
problems, with the CIBVP standing for the Bondi-Sachs gauge used in
characteristic evolutions. We wish to understand whether or not
problems with these features can be successfully used for CCE or CCM
in principle.

For the IBVP estimate our starting point is
\begin{align*}
  \p_t ||\mathbf{\bar{u}}||^2_{L^2(\Sigma_t)} =
  \p_t \int_{\Sigma_t} \mathbf{\bar{u}}^T \mathbf{\bar{u}} =
  \p_t \int_{\Sigma_t} \left(\bar{\phi}^2 + \bar{\psi_\varv}^2
  + \bar{\psi}^2 \right)
  \,,
\end{align*}
which after using~\eqref{eqn:CCM_IBVP_sys}, the divergence theorem
assuming~$\mathbf{\bar{u}} \rightarrow \mathbf{0}$
as~$\rho \rightarrow - \infty$ and integrating in the~$t$ domain,
yields
\begin{align}
  ||\mathbf{\bar{u}}||^2_{L^2(\Sigma_{t_f})}
  + ||\mathbf{\bar{u}}||^2_{L^2_\textrm{out} (\mathcal{T}_0)}
  =
  ||\mathbf{\bar{u}}||^2_{L^2(\Sigma_0)}
  + ||\mathbf{\bar{u}}||^2_{L^2_\textrm{in} (\mathcal{T}_0)}
  \,,
  \label{eqn:IBVP_estimate}
\end{align}
where~$||\mathbf{\bar{u}}||^2_{L^2_\textrm{out}(\mathcal{T}_0)}$
denotes integral over~$\mathcal{T}_0$ that contains only the outgoing
fields~$\bar{\phi},\,\bar{\psi_\varv}$, and similarly for the
ingoing. The estimate~\eqref{eqn:IBVP_estimate} states that the energy
of the solution equals the energy of its given data, so that the
solution is controlled by the given data.

In a Cauchy-type setup we specify all fields on the initial spacelike
hypersurface and, by solving the system we obtain all of them on
spacelike hypersurfaces to the future. On the contrary, in a
single-null characteristic setup, fields with ``evolution'' equations
are chosen on the initial null hypersurface and those that satisfy
equations intrinsic to the null hypersurfaces are specified as
boundary data. As we will see in the following, this has a natural
consequence on the type of estimates that we can hope to demonstrate,
both in terms of the domain on which we integrate and the particular
fields that appear. This is due to the geometry of the setup.

Motivated from the IVP estimates in Fourier space of
subsection~\ref{Subsubsection:homogeneous_WH}
and\ref{Subsubsection:inhomogeneous_WH} we might naively first
consider for the CIBVP the \textit{lopsided norm}
\begin{align*}
  ||\mathbf{u}||^2_{q (\mathcal{D})} = \int_{\mathcal{D}}
  \left(
  \phi^2 + \psi_\varv^2 + \psi^2 + \left( \p_z \phi \right)^2
  \right)
  \,,
\end{align*}
in some domain~$\mathcal{D}$, where only~$\p_z \phi$ is added to the
integrand of the~$L^2$-norm, because precisely this term causes the
pathological structure in the angular principal part of the WH
model. Following our previous discussion however, it is more
appropriate to split the integrand into separate pieces for the
ingoing and outgoing variables. The domain~$\mathcal{D}$
becomes~$\mathcal{N}_{u}$ and $\mathcal{T}_{x}$ respectively for
each. For the ingoing variables we start from
\begin{align*}
  \p_u ||\mathbf{u}||^2 _{q_\textrm{in}(\mathcal{N}_u)} =
  \p_u \int_{\mathcal{N}_u} \psi^2
  \,,
\end{align*}
since there are no~$\p_u$ equations for the outgoing ones. We assume
that~$\psi \rightarrow 0$ as~$x \rightarrow x_f$ in the given data,
which is the analog in our model to requiring no incoming
gravitational waves from future null infinity, working on a
compactified radial domain. After using~\eqref{eqn:CCM_CIBVP_sys_c},
the divergence theorem and integrating in the~$u$ domain we obtain
\begin{align}
  2 ||\mathbf{u}||^2 _{q_\textrm{in}(\mathcal{N}_{u_f})} +
  ||\mathbf{u}||^2 _{q_\textrm{in}(\mathcal{T}_0)}
  =
  2 ||\mathbf{u}||^2 _{q_\textrm{in}(\mathcal{N}_0)}
  \,.
  \label{eqn:CIBVP_in_estimate}
\end{align}
For the outgoing variables the starting point is
\begin{align*}
  \p_x ||\mathbf{u}||^2_{q_\textrm{out}(\mathcal{T}_x)} =
  \p_x \int_{\mathcal{T}_x}
  \left(
  \phi^2 + \psi_\varv^2 +
  \left( \p_z \phi \right)^2
  \right)
  \,,
\end{align*}
and by using~\eqref{eqn:CCM_CIBVP_sys_a} and \eqref{eqn:CCM_CIBVP_sys_b}, the
divergence theorem and integrating in the~$x$ domain up to some
arbitrary~$x^\prime$ we obtain
\begin{align}
  ||\mathbf{u}||^2_{q_\textrm{out}(\mathcal{T}_{x^\prime})}
  =
  ||\mathbf{u}||^2_{q_\textrm{out}(\mathcal{T}_0)}
  +
  \int_0^{x^\prime}
  \left(
  \int_{\mathcal{T}_x} 2 \psi_\varv \p_z \phi
  \right)
  \, dx
  \,,
  \label{eqn:CIBVP_out_estimate_1}
\end{align}
where the last term is due to the hyperbolicity of the system and
would not appear for our SH example. Using
$ 2 \psi_\varv \p_z \phi \leq \phi^2 + \psi_\varv^2 + \left( \p_z \phi
\right)^2$ the latter reads
\begin{align*}
  ||\mathbf{u}||^2_{q_\textrm{out}(\mathcal{T}_{x^\prime})}
  \leq
  ||\mathbf{u}||^2_{q_\textrm{out}(\mathcal{T}_0)}
  +
  \int_0^{x^\prime} ||\mathbf{u}||^2_{q_\textrm{out}(\mathcal{T}_x)} \, dx
  \,,
\end{align*}
and by applying Gr\"onwall's inequality we obtain
\begin{align}
  ||\mathbf{u}||^2_{q_\textrm{out}(\mathcal{T}_{x^\prime})}
  \leq
  e^{x^\prime}
  ||\mathbf{u}||^2_{q_\textrm{out}(\mathcal{T}_0)}
  \,.
  \label{eqn:CIBVP_out_estimate_2}
\end{align}
Hence, the energy of the outgoing fields at each arbitrary timelike
hypersurface~$\mathcal{T}_{x^\prime}$ in the characteristic domain is
bounded. The sum of~\ref{eqn:CIBVP_in_estimate}
and~~\ref{eqn:CIBVP_out_estimate_2} is the complete energy estimate
for the CIBVP and yields
\begin{equation}
  \begin{aligned}
    &
    2||\mathbf{u}||^2 _{q_\textrm{in}(\mathcal{N}_{u_f})} +
    ||\mathbf{u}||^2 _{q_\textrm{in}(\mathcal{T}_0)} +
    \textrm{sup}_{x'} ||\mathbf{u}||^2_{q_\textrm{out}(\mathcal{T}_{x^\prime})}
    \\
    & \qquad \qquad \quad
    \leq
    2 ||\mathbf{u}||^2 _{q_\textrm{in}(\mathcal{N}_0)}
    +
    e^{x_f}
    ||\mathbf{u}||^2_{q_\textrm{out}(\mathcal{T}_0)}
    \,,
  \end{aligned}
  \label{eqn:CIBVP_estimate_full}
\end{equation}
where we used that~$e^{x^\prime} \leq e^{x_f}$ for~$x^\prime \in
[0,x_f]$ and chose the supremum
of~$||\mathbf{u}||^2_{q_\textrm{out}(\mathcal{T}_{x^\prime})}$ to
obtain the largest possible bounded left hand side, since the outgoing
lopsided norm is not necessarily monotonically increasing with~$x$.
Thus, the energy of the solution to the CIBVP is controlled by the
given data on~$\mathcal{N}_0$ and~$\mathcal{T}_0$.

We first interpret these estimates in the framework of CCE. Choosing
suitable data, our estimate for the IBVP shows that one obtains a
smooth solution in the domain of the Cauchy-type setup. One can then
use this solution to provide boundary data on~$\mathcal{T}_0$ for the
CIBVP that are finite also in the lopsided norm, and the solution to
this characteristic problem has a good energy estimate as shown
earlier too. Hence the CCE process is perfectly valid for our model,
and provided analogous estimates for GR in the Bondi-like gauges used,
would be in that context too. One question that arises for GR, but
which for now we have no insight, is whether or not this procedure
excludes any data of interest. For CCM the discussion is rather
different, since IBVP and CIBVP are solved simultaneously and data are
communicated between domains. Effectively, one joins the PDE problems
and they may be viewed as one. Hence, let us try to obtain an energy
estimate for the joint PDE problem, by
adding~\eqref{eqn:IBVP_estimate} and~\eqref{eqn:CIBVP_estimate_full}:
\begin{equation}
  \begin{aligned}
    &
    ||\mathbf{u}||^2_{L^2(\Sigma_{t_f})} +
    ||\mathbf{u}||^2_{L^2_\textrm{out} (\mathcal{T}_0)} +
    2 ||\mathbf{u}||^2 _{q_\textrm{in}(\mathcal{N}_{u_f})} +
    \\
    &
    ||\mathbf{u}||^2 _{q_\textrm{in}(\mathcal{T}_0)} +
    \textrm{sup}_{x'} ||\mathbf{u}||^2_{q_\textrm{out}(\mathcal{T}_{x^\prime})}
  \\
  & \qquad \qquad \quad
    \leq \,
    ||\mathbf{u}||^2_{L^2(\Sigma_0)} +
    ||\mathbf{u}||^2_{L^2_\textrm{in} (\mathcal{T}_0)} +
    \\
    &
    \qquad \qquad \qquad
    2 ||\mathbf{u}||^2 _{q_\textrm{in}(\mathcal{N}_0)}
    +
    e^{x_f}
    ||\mathbf{u}||^2_{q_\textrm{out}(\mathcal{T}_0)}
    \,,
  \end{aligned}
  \label{CCM_estimate}
\end{equation}
where now~$\mathbf{\bar{u}} = \mathbf{u}$. For the joint problem there
is `effectively' no boundary~$\mathcal{T}_0$ at which we are free to
choose data, and hence any estimate should not involve integrals over
this domain. The relevant terms can however cancel each other only if
the two norms that appear coincide. This requires either that the
CIBVP relies on a symmetric hyperbolic PDE system and hence is
well-posed in the~$L^2$-norm (see for instance~\cite{BisGomHol96,
  BisGomHol97, Cal06}), or that the IBVP relies on a system that is
well-posed in the same lopsided norm as the CIBVP. But this requires
special structure, above and beyond symmetric hyperbolicity, on the
equations used in the IBVP. Regarding GR, the first option would
translate into developing a SH (hopefully also symmetric hyperbolic)
single-null formulation and the second to building a formulation that
is well-posed in the same lopsided norm that Bondi-like gauges
(perhaps) are. Given the long search for formulations that {\it work}
for practical evolution however, such an artisanal construction seems
poorly motivated. In summary; unless special structure is present in
the field equations solved for the IBVP, the solution to the weakly
hyperbolic CIBVP cannot be combined with that of an IBVP of a
symmetric hyperbolic system in such a way as to provide a solution to
the whole problem which has an energy bounded by that of the given
data.

\section{Numerical Experiments} \label{Section:numerical_experiments}

We now use the toy models introduced in Sec. \ref{Section:toy_models}
to diagnose the effects of weak hyperbolicity at the numerical
level. We perform convergence tests in the single-null setup for both
the WH and SH models in a discrete approximation to the~$L^2$-norm,
for smooth and noisy given data. We also perform convergence tests
with noisy given data in the lopsided norm, for the different versions
of the WH model analyzed in the previous section.

\subsection{Implementation} \label{Subsection:numerical_implementation}

As in other schemes to solve the CIBVP, several different ingredients
are needed in the algorithm. These can be summarized for our
models~\eqref{eqn:WH_model} and~\eqref{eqn:SH_model} as follows:
\begin{enumerate}

\item The domain of the PDE problem is~$x \in [0,1]$,~$z \in[0,2\pi)$
  with periodic boundary conditions and~$u \in[u_0, u_f]$, with~$u_0$
  and~$u_f$ the initial and final times respectively. We always
  include the point~$x=1$ in the computational domain so that we do
  not need to impose boundary conditions at the outer boundary, since
  there are no incoming characteristic variables there.

\item For the initial time~$u_0$ provide initial data~$\psi(u_0,x,z)$
  on the surface~$u=u_0$ and boundary data~$\phi(u_0,0,z)$
  and~$\psi_\varv(u_0,0,z)$.

\item Integrate the intrinsic equations of each model to
  obtain~$\phi(u_0,x,z)$ and~$\psi_\varv(u_0,x,z)$. We perform this
  integration using the two-stage, second order strong stability
  preserving method of Shu and Osher (SSPRK22)~\cite{ShuOsh88}.
  
\item Integrate the evolution equation of each model to
  obtain~$\psi(u_1,x,z)$ at the surface~$u=u_1=u_0 + \Delta u$. We
  choose~$\Delta u = 0.25 \Delta x$ to satisfy the
  Courant-Friedrichs-Lewy (CFL) condition and the numerical
  integration is performed using the fourth order Runge-Kutta (RK4)
  method.

\item Any derivative appearing in the right-hand-sides of these
  integrations is approximated using second order accurate centered
  finite difference operators, except at the boundaries, where second
  order accurate forward and backward difference operators are used
  respectively. 
  
\item Providing boundary data~$\phi(u,0,z)$ and~$\psi_\varv(u,0,z)$ as
  in the PDE specification~\eqref{eqn:model_boundary_data}, we repeat
  steps~$2$ and~$3$ to obtain~$\phi(u,x,z)$,~$\psi_\varv(u,x,z)$
  and~$\psi(u,x,z)$ until the final time~$u_f$. This is the solution
  of the PDE.
  
\end{enumerate}
No artificial dissipation is introduced. The implementation was made
using the Julia language~\cite{BezEdeKar17} with the
DifferentialEquations.jl package~\cite{RacNie17} to integrate the
equations. Our code is freely available~\cite{GiaHilZil20_public}. We
apply convergence tests to our numerical scheme for both toy
models. The tests are performed for smooth, as well as for noisy given
data. The latter are often called robust stability tests. They form
part of the Mexico-city testbed for numerical
relativity~\cite{AlcAllBau03}. These tests have been performed widely
in the literature~\cite{CalHinHus05,Hin05,BoyLinPfe06,BabHusHin07,
  WitHilSpe10, CaoHil11}, often, as in our case, with adaptations for
the setup under consideration.

\subsection{Convergence tests} \label{Subsection:convergence_tests}

By~\textit{convergence} we mean the requirement that the difference
between the numerical approximation provided by a finite difference
scheme and the exact solution of the continuum PDE system tends to
zero as the grid spacing is increased. The finite difference scheme is
called~\textit{consistent} when it approximates the correct PDE system
and the degree to which this is achieved is its~\textit{accuracy}. The
scheme is called~\textit{stable} if it satisfies a discretized version
of~\eqref{wp_inequality} or~\eqref{weak_wp_inequality}. In this
context versions of each continuum norm is replaced by a suitable
discrete analog. Here we replace the~$L^2$-norm for the single-null
setup with
\begin{equation}
  \begin{aligned}
    &
    ||\mathbf{u}||_{h_u,h_x,h_z}^2 =
    \sum_{x,z} \, \psi^2 \, h_x \, h_z
    +
    \\
    & \qquad \qquad \qquad
    \textrm{max}_x \sum_{u,z} \, \left(\phi^2 + \psi_\varv^2  \right)
    h_u \, h_z  \,,
  \end{aligned}
  \label{discrete_L2}
\end{equation}
with the first sum taken over all points on the grid, with~$h_x$
and~$h_z$ the grid-spacing in the~$x$ and~$z$ directions respectively,
and the second sum over all points in the~$z$ and~$u$ directions
($h_u = 0.25 h_x$ for our setup), for all~$x$ grid points and keeping
the maximum in the $x$ direction. The first sum involves only ingoing
and the second only outgoing variables. When, as will be the case in
what follows, we have~$h_x=h_z=h$ we label the norm simply
with~$h$. Our discrete approximation to the lopsided norm is,
\begin{equation}
  \begin{aligned}
    &
    ||\mathbf{u}||^2_{\textrm{q}(h_u, h_x , h_z)}
    =
    \sum_{x,z} \, \psi^2 \, h_x \, h_z
    +
    \\
    & \qquad \qquad
    \textrm{max}_x \sum_{u,z} \, \left(\phi^2 + \psi_\varv^2 +
      \left(D_z \phi\right)^2
    \right)
    h_u \, h_z  \,,
  \end{aligned}
  \label{discrete_lopsided}
\end{equation}
where~$D_z$ is the second order accurate, centered, finite difference
operator that replaces the continuum operator~$\p_z$, by
\begin{align}
  D_z f_{h}(x_i) = \frac{f_{h}(x_{i+1}) - f_{h}(x_{i-1})}{2h_z}
  \, ,\label{FT_stencil}
\end{align}
for a grid function~$f_h$ on a grid with spacing~$h_z$. When the two
grid spacings are equal we again label the norm simply with~$h$. This
approximation to the continuum lopsided norm is not unique. If we were
attempting to prove that a particular discretization converged, it
might be necessary to take another. Denoting by~$f$ the solution to
the continuum system and as~$f_h$ the numerical approximation at
resolution~$h$ provided by a convergent finite difference scheme of
accuracy~$n$, then
\begin{align}
  f = f_h + O\left( h^n\right) \,,
  \label{numerical_approx}
\end{align}
and hence
\begin{align}
  ||f - f_h|| = O(h^n)\, ,
  \label{exact_convergence_relation}
\end{align}
in some appropriate norm~$|| \cdot ||$ on the grid, with the
understanding that the exact solution should be evaluated on said
grid. Full definitions of the notions of consistency, stability and
convergence for the IVP can be found, for example,
in~\cite{GusKreOli95,Tho98c,Hin05}.

We use a second order accurate numerical approximation, so
that~$n=2$. Considering numerical evolutions with coarse, medium and
fine grid spacings~$h_c$,~$h_m$ and~$h_f$ respectively, we can
construct a useful quantity for these tests
\begin{align}
  Q \equiv \frac{h_c^n -h_m^n}{h_m^n - h_f^n}
  \,, \label{convergence_factor}
\end{align}
which we call~\textit{convergence factor}. In our convergence tests we
solve the same discretized PDE problem for different resolutions and
every time we want to increase resolution we halve the grid-spacing in
all directions i.e.
\begin{align*}
  h_m = h_c/2 \,, \quad h_f = h_c/4 \,.
\end{align*}
Following this approach the convergence factor
is~$Q=4$. Combining~\eqref{numerical_approx}
and~\eqref{convergence_factor} one can obtain the relation
\begin{align}
  f_{h_c} - f_{h_c/2} = Q \left(f_{h_c/2} - f_{h_c/4} \right)
  \,, \label{pointwise_conv_relation}
\end{align}
understood on shared grid-points in the obvious way, which is used to
investigate pointwise convergence. In what follows the different
resolutions are denoted as
\begin{align*}
  h_q = h_0/2^q \,.
\end{align*}
The lowest resolution~$h_0$ has $N_x=17$ points in the $x$-grid
and~$N_z=16$ in the $z$-grid. We work in units of the code in the
entire section.

\subsubsection{Smooth data} \label{Subsubsection:smooth_data}

For the simulations with smooth given data the initial and final times
are~$u_0=0$ and~$u_f=1$ respectively. For both toy models we provide as
initial data
\begin{align*}
  \psi(0,x,z)=e^{-100 \left( x-1/2 \right)^2} \, \sin(z)\,,
\end{align*}
and as boundary data
\begin{align*}
\phi(u,0,z)= 3 \, e^{-100 \left( u-1/2 \right)^2} \, \sin(z)\,,
\end{align*}
and
\begin{align*}
\psi_\varv(u,0,z)=e^{-100 \left( u-1/2 \right)^2} \,\sin(z)\,.
\end{align*}
For the SH model we choose the following source terms
\begin{align}
  -S_\phi = \psi \,,
  \quad -S_{\psi_\varv} = \phi + \psi \,,
  \quad -S_\psi = \phi\,,
\end{align}
and for the WH model we choose the homogeneous case. As discussed in
Sec.~\ref{Subsection:algebraic_characterization}, well-posedness of
the SH model is unaffected by lower order source terms, so the
specific choice of source terms here is not vital. However, we choose
to work with the homogeneous WH model, because weakly well-posed
problems are sensitive to lower order perturbations.

\begin{figure}[t]
  \subfigure[\,SH~$\phi(1,x,z)$.]
  {\includegraphics[width=0.23\textwidth,height=3cm]{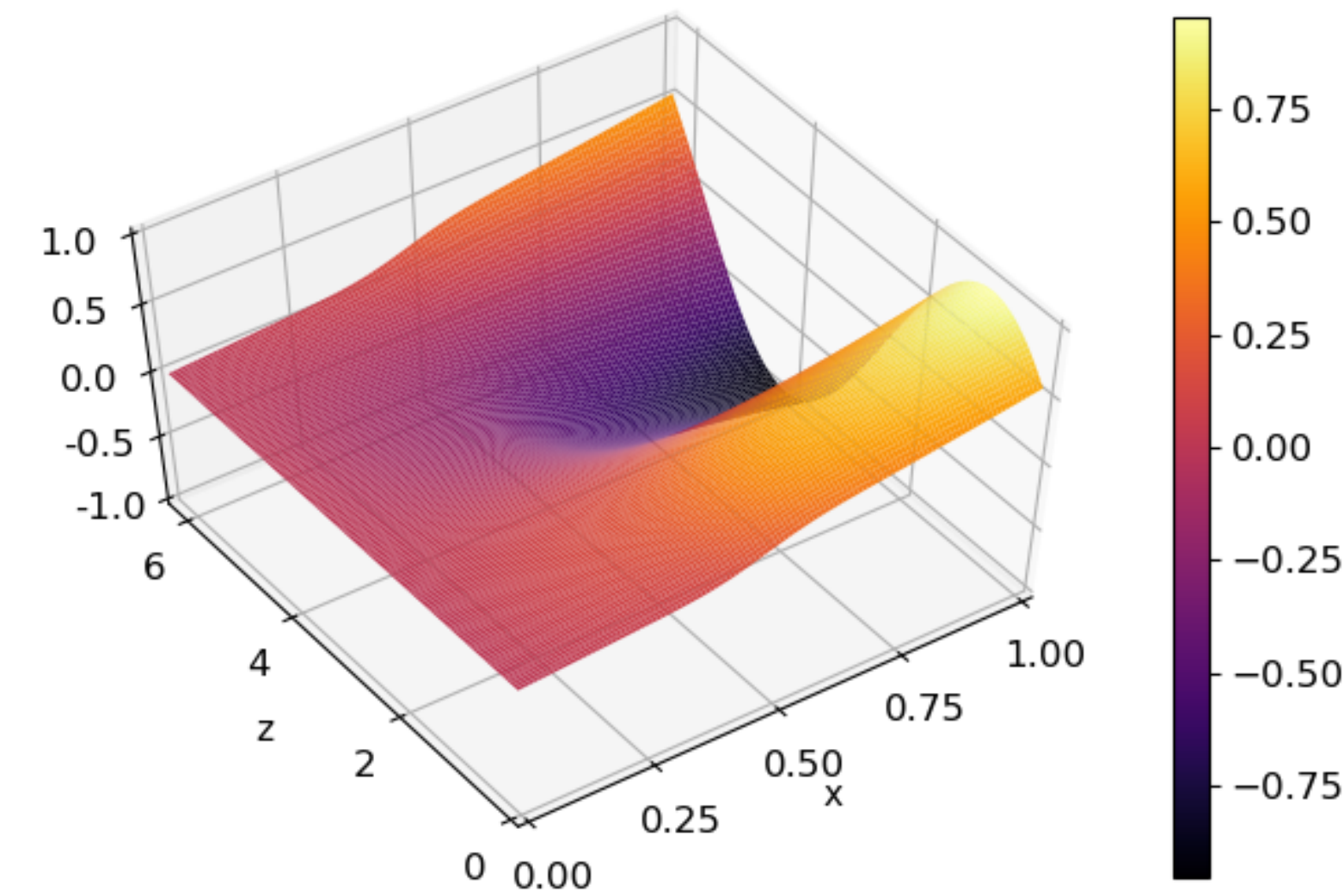}}
  \subfigure[\,WH~$\phi(1,x,z)$.]
  {\includegraphics[width=0.23\textwidth,height=3cm]{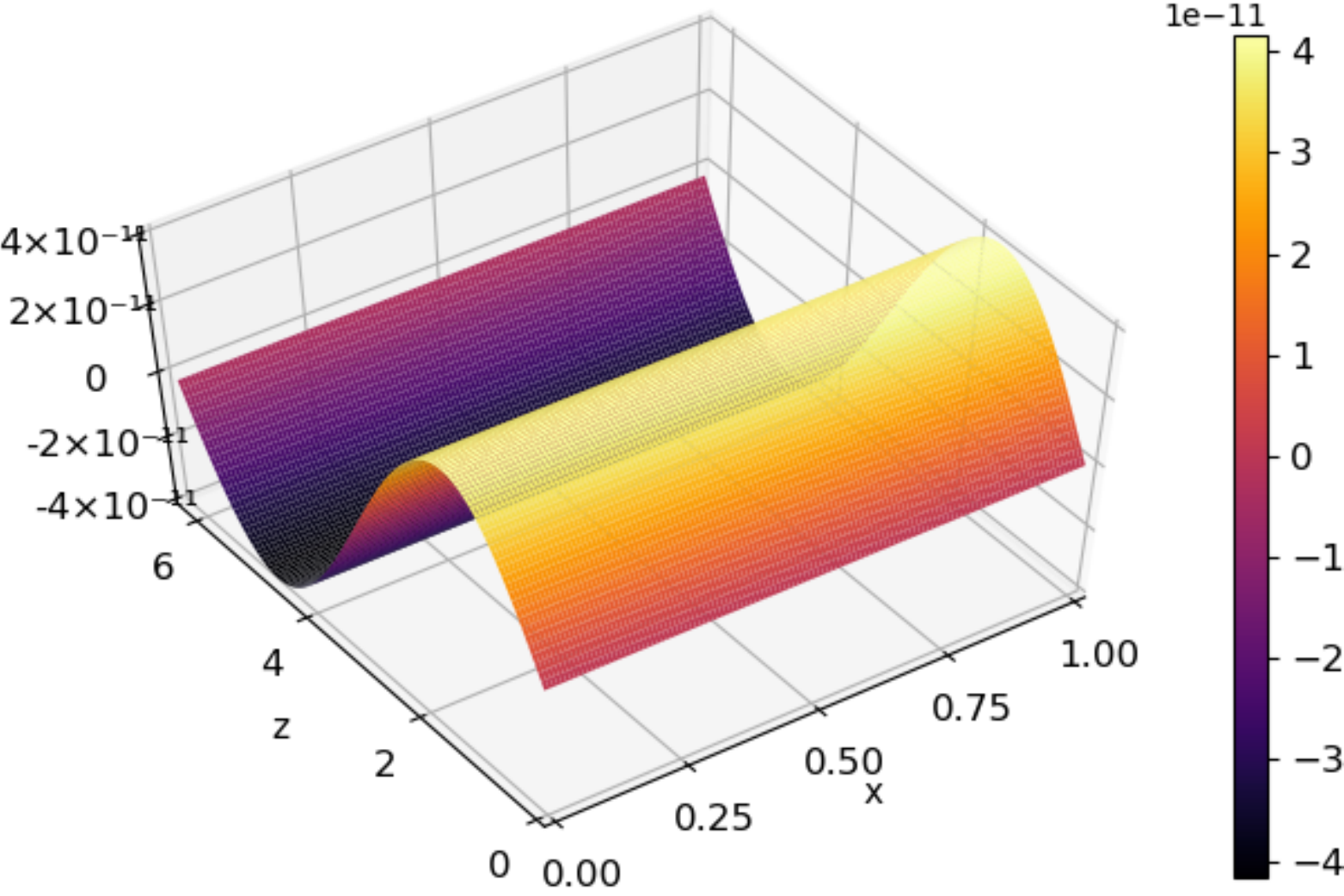}}
  \subfigure[\,SH~$\psi_\varv(1,x,z)$.]
  {\includegraphics[width=0.23\textwidth,height=3cm]{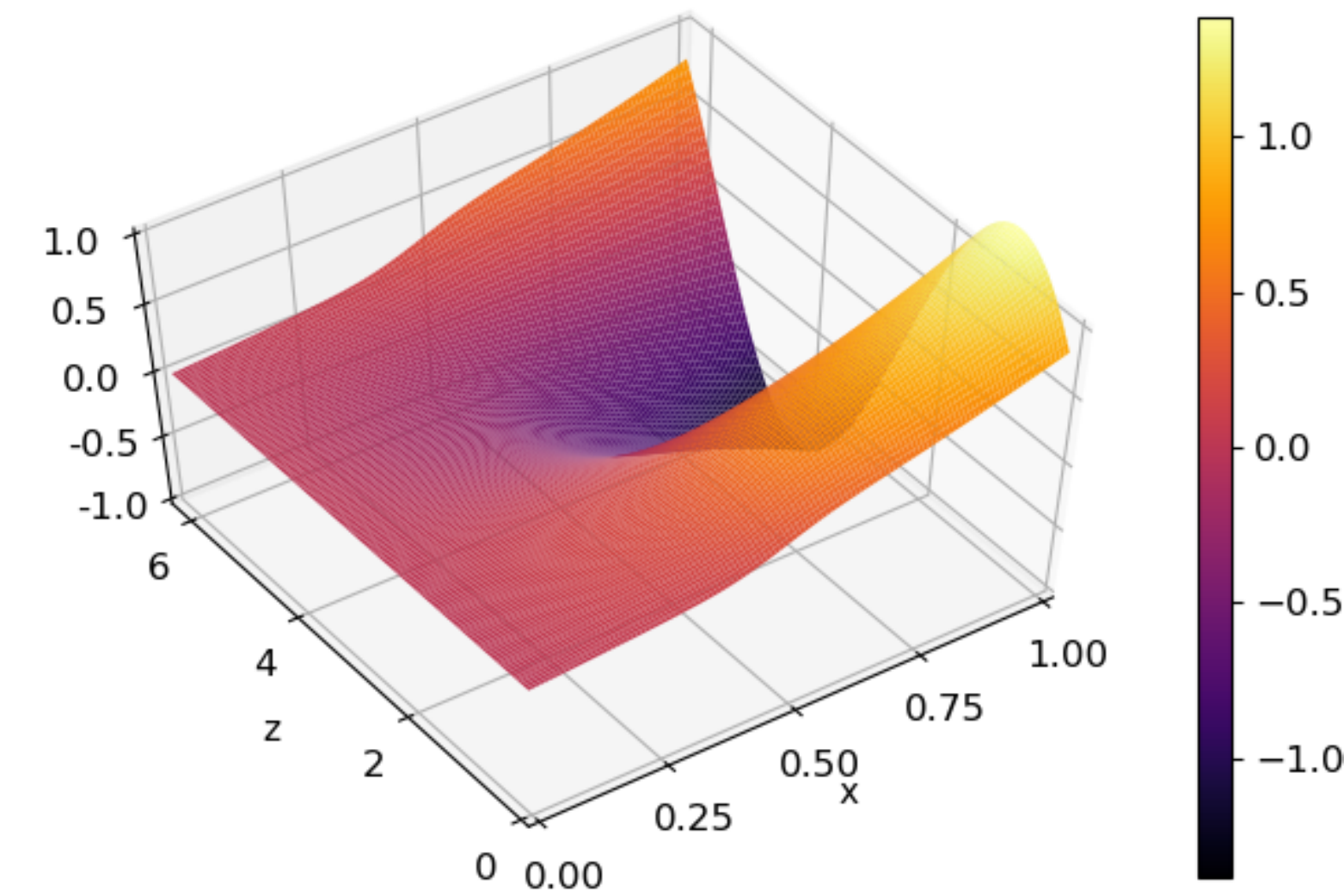}}
  \subfigure[\,WH~$\psi_\varv(1,x,z)$.]
  {\includegraphics[width=0.23\textwidth,height=3cm]{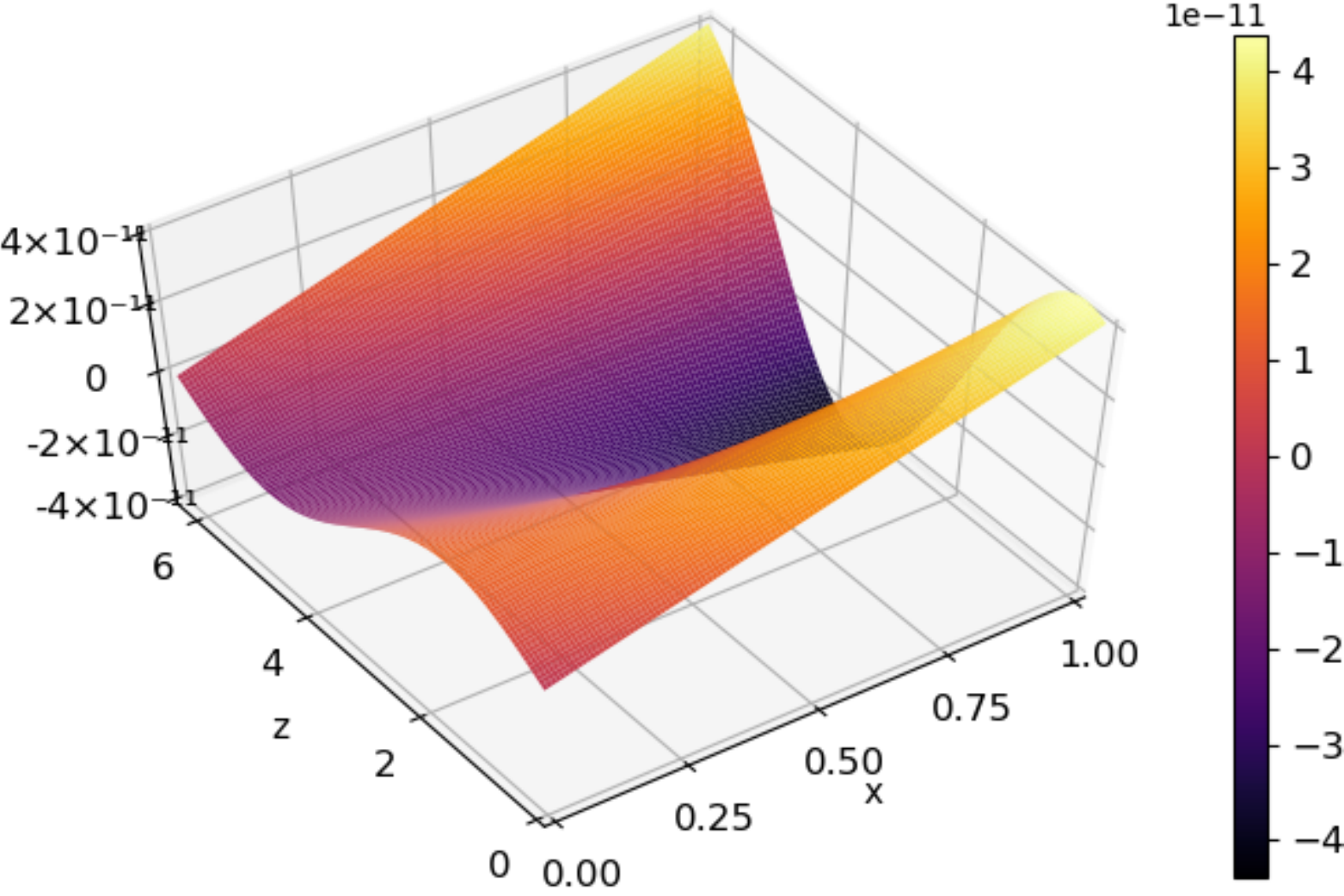}}
  \subfigure[\,SH~$\psi(1,x,z)$.]
  {\includegraphics[width=0.23\textwidth,height=3cm]{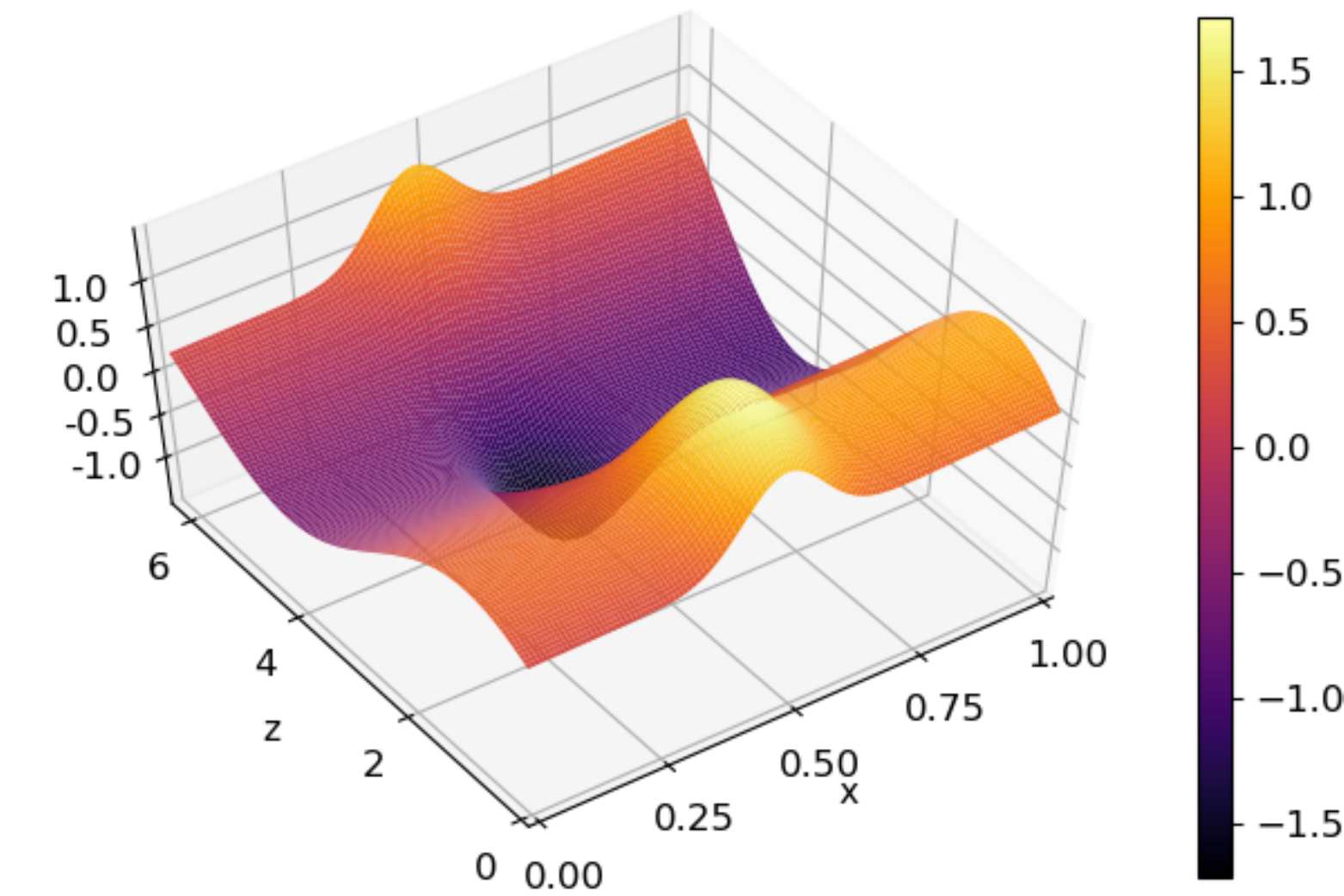}}
  \subfigure[\,WH~$\psi(1,x,z)$.]
  {\includegraphics[width=0.23\textwidth,height=3cm]{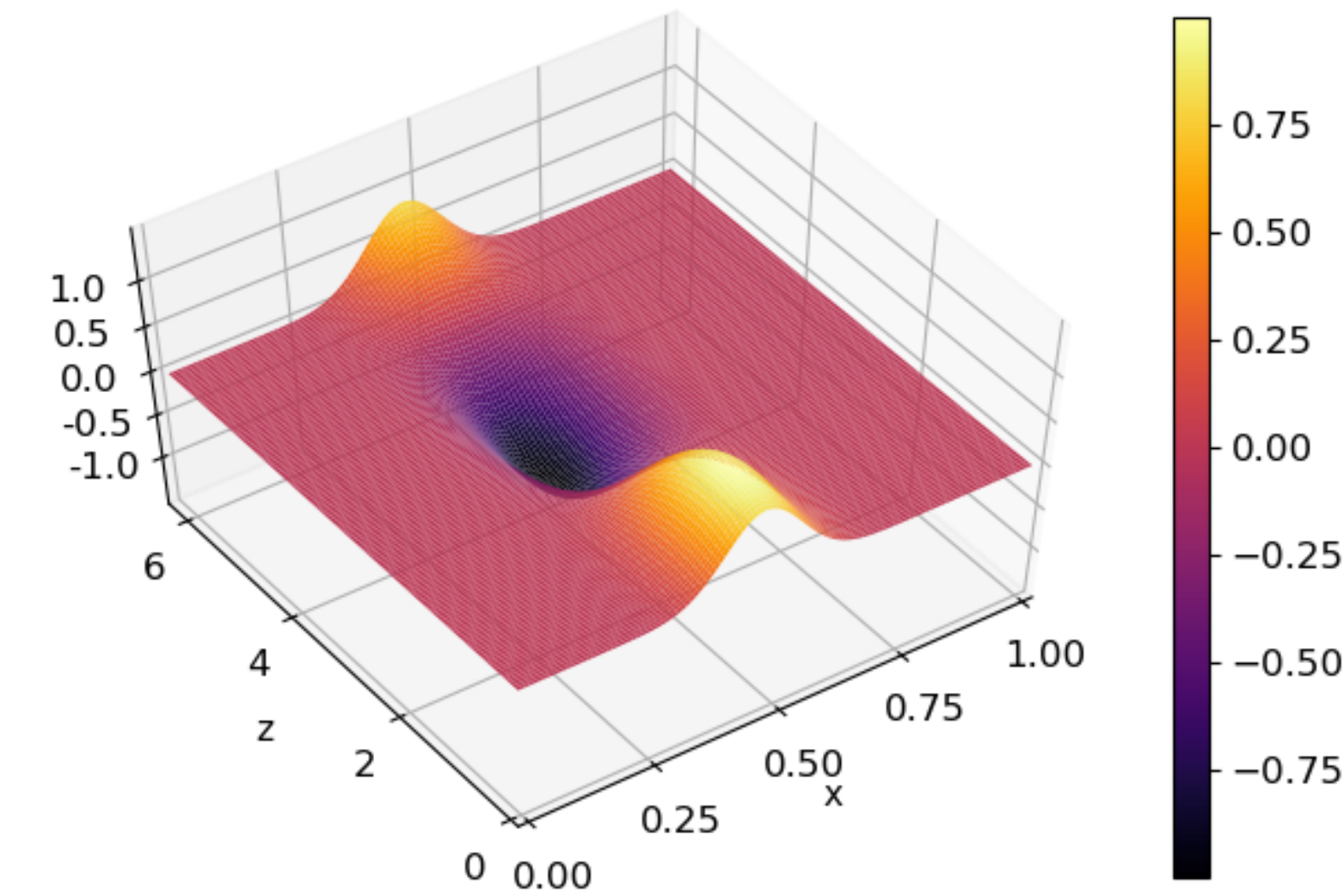}}
  \caption{The fields~$\phi$, ~$\psi_\varv$ and~$\psi$ at final
    evolution time~$u = 1$, for the SH model (left) and the
    homogeneous WH model (right), with the same smooth given data.
    Observe that the fields~$\phi$ and~$\psi_\varv$ in the WH case are
    still of the same magnitude~$\sim10^{-11}$ as the boundary data at
    the retarded time~$u=1$. This is not true once generic source
    terms are taken. \label{Fig:smooth_sols}}
\end{figure}

\begin{figure}[t]
  \includegraphics[width=0.5\textwidth]{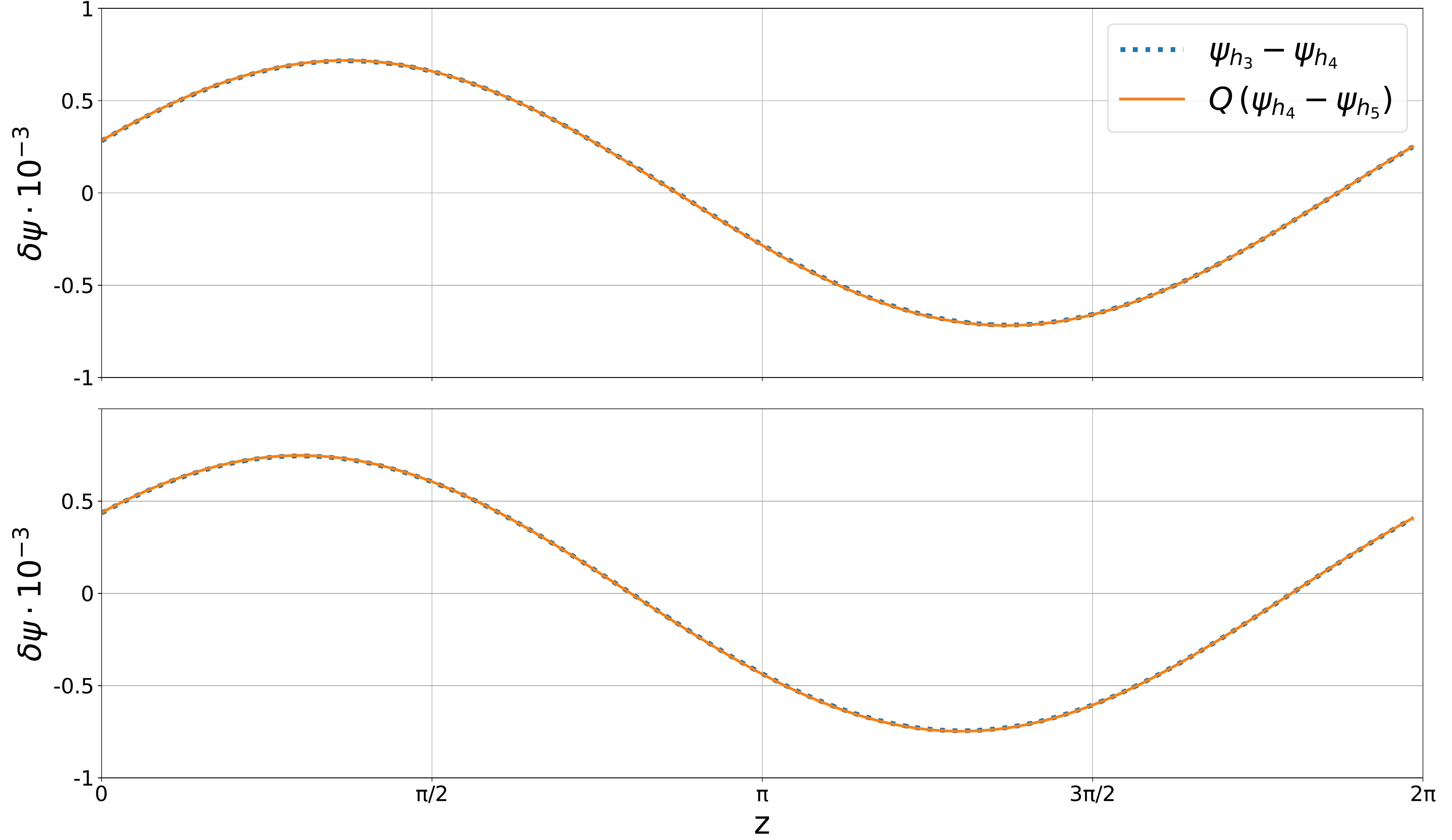}
  \centering{
    \caption{Here we plot simultaneously~$\psi_{h_3} - \psi_{h_4}$
      and~$Q \left( \psi_{h_4} - \psi_{h_5}\right)$, for the SH (top)
      and the WH (bottom) toy models. We fix~$x=0.5$. Since our scheme
      is second order and we are doubling resolution we fix~$Q=4$. The
      results for fixed~$z$ are similar. The plot is compatible with
      perfect second order pointwise convergence.}
  \label{Fig:pointwise_convergence}}
\end{figure}

Runs with resolutions~$h_0,\, h_1,\, h_2 ,\, h_3,\, h_4$ and $h_5$
were performed. In Fig.~\ref{Fig:smooth_sols} the basic dynamics are
plotted with each model. To first verify that the numerical scheme is
implemented successfully we performed pointwise convergence tests for
both models. We focus specifically here on the highest three
resolutions. The algorithm is the following:
\begin{enumerate}

\item Consider~$h_3$, $h_4$ and~$h_5$ as coarse, medium and fine
  resolutions, respectively.
  
\item Calculate $\psi_{h_3} - \psi_{h_4}$ and
  $\psi_{h_4} - \psi_{h_5}$ for the gridpoints of $h_3$, for the final
  timestep of the evolution.

\item Plot simultaneously $\psi_{h_3} - \psi_{h_4}$ and
  $Q \left( \psi_{h_4} - \psi_{h_5}\right)$. As indicated
  from~\eqref{pointwise_conv_relation}, for a convergent numerical
  scheme the two quantities should overlap, when multiplying the
  latter with the appropriate convergence factor.
  
\end{enumerate}
In Fig.~\ref{Fig:pointwise_convergence} we illustrate the results of
this test for the aforementioned smooth given data for both models. At
this resolution one clearly observes perfect pointwise convergence in
both cases.

\begin{figure*}[t]
  \includegraphics[width=1.\textwidth]{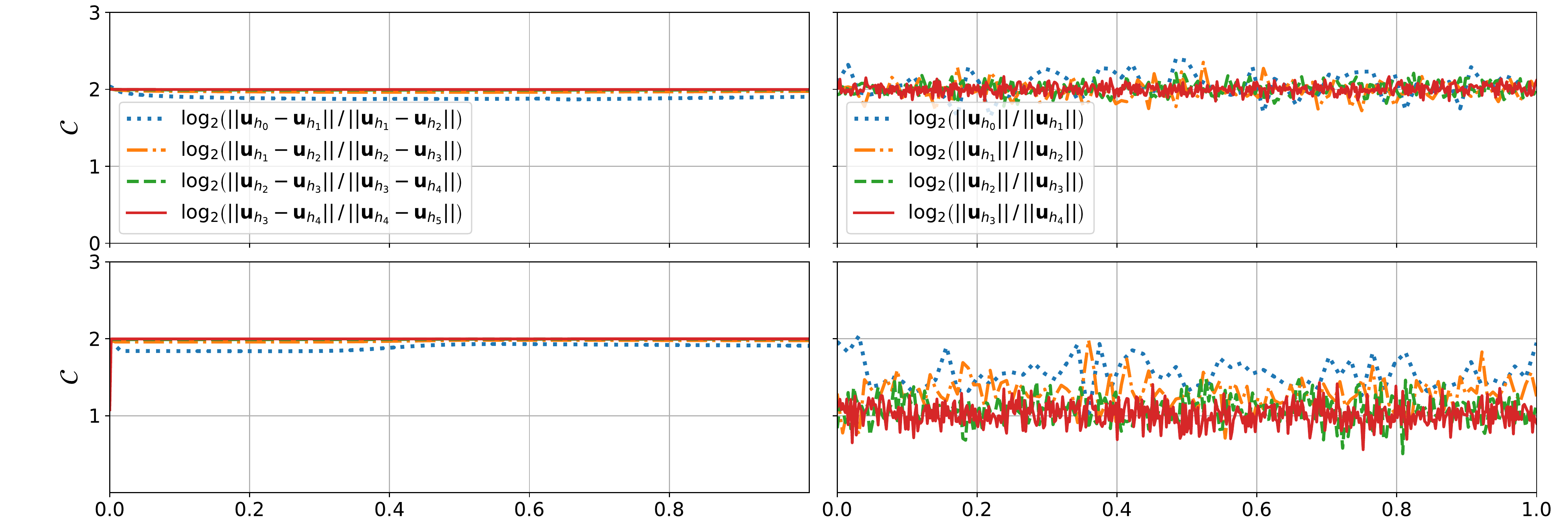}
  \centering{
    \caption{The convergence ratio in the~$L^2$-norm, for the strongly
      (above) and the weakly (below) toy models, for smooth (left) and
      noisy (right) given data, as a function of the simulation
      time. All plots have the same scale on the~$y$-axis. For smooth
      given data we consider the self convergence
      ratio~\eqref{self_conv_ratio_log2} while for noisy given data
      the exact convergence ration~\eqref{exact_conv_ratio}. If we
      consider the self convergence ratio also for the noisy case the
      results are qualitatively the same.\label{fig:L2_norms_all}}}
\end{figure*}

We also wish to examine convergence of our numerical solutions in
discrete approximations of the aforementioned norms. Given that the
exact solution to the PDE problem is unknown and that each time we
increase resolution we decrease the grid spacing in all directions by
a factor of~$d$, we can build the following useful quantity
\begin{align}
  \mathcal{C}_\textrm{self} =
  \log_d
  \frac{|| \mathbf{u}_{h_c} - \perp^{h_c/d}_{h_c}\mathbf{u}_{h_c/d} ||_{h_c} }
       { || \perp^{h_c/d}_{h_c} \mathbf{u}_{h_c/d}
         - \perp^{h_c/d^2}_{h_c}\mathbf{u}_{h_c/d^2} ||_{h_c} }\, ,
       \label{self_conv_ratio}
\end{align}
which we call~\textit{self-convergence ratio}, with
~$\mathbf{u}=\left(\phi, \psi_\varv, \psi \right)^T$ the state vector
of the PDE system and~$\phi$,~$\psi_\varv$,~$\psi$ grid
functions. Here~$\perp^{h_c/d}_{h_c}$ denotes the projection (in our
setup injection) operator from the~$h_c/d$ grid onto the~$h_c$ grid.
We calculate~$\mathcal{C}_\textrm{self}$ for a discrete analog of
the~$L^2$-norm. However, if one wishes to examine convergence in a
different norm, $L^2$ can be replaced with that. The theoretical value
of~$\mathcal{C}_\textrm{self}$ equals the accuracy~$n$ of the
numerical scheme, and in our specific setup
\begin{align}
  \mathcal{C}_\textrm{self} = \textrm{log}_2 \frac{|| \mathbf{u}_{h_c}
    - \perp^{h_c/2}_{h_c}\mathbf{u}_{h_c/2} ||_{h_c} } { ||
    \perp^{h_c/2}_{h_c}\mathbf{u}_{h_c/2} -
    \perp^{h_c/4}_{h_c}\mathbf{u}_{h_c/4} ||_{h_c} } = 2 \,.
  \label{self_conv_ratio_log2}
\end{align}
We obtain numerical solutions for the same smooth given data for both
models at the various resolutions mentioned before. For triple of
resolution, double resolution and quadruple resolution, we project all
gridfunctions onto the coarse grid, and compute~$C_\textrm{self}$ at
its timesteps. In the left panel of Fig.~\ref{fig:L2_norms_all} we
collect the results of these norm convergence tests. Both models show
similar behavior. At low resolutions curve drifts from the desired
rate at early times, but the situation improves as we increase
resolution, with~$C_\textrm{self}$ approaching the expected value. The
trend with increasing resolution is the essential behavior we are
looking at in these tests. By limiting ourselves to convergence tests
with smooth given data we could be misled that the WH toy model
provides a well-posed CIBVP in the~$L^2$-norm, since the numerical
solutions appear to converge in this norm during our simulations. In
other words, were we ignorant of the hyperbolicity of the system, it
would be impossible to distinguish strongly and weakly hyperbolic PDEs
with this test.
 
\subsubsection{Noisy data}

One can also perform norm convergence tests with random noise as given
data, which is a strategy to simulate numerical error in an
exaggerated form. Since it is expected that numerical error decreases
as resolution increases, when performing simulations for these tests
one must scale appropriately the amplitude of the noise as resolution
improves. This scaling is important to construct a sequence of initial
data that converges in a suitable norm to initial data appropriate for
the continuum system. The choice of norm here is essential, and should
be one which, if possible, provides a bound for the solution of a
(weakly) well-posed PDE problem, in the sense of~\eqref{wp_inequality}
and~\eqref{weak_wp_inequality}.

For these tests we perform simulations where the smooth part of the
given data is trivial (zero), and hence the exact solution for every
PDE problem based on our models vanishes identically. Knowing the
exact solution, in addition to the self convergence
rate~\eqref{self_conv_ratio}, we can also construct the exact
convergence ratio
\begin{align}
  \mathcal{C}_\textrm{exact} = \log_d
  \frac{|| \mathbf{u}_{h_c} - \mathbf{u}_\textrm{exact} ||_{h_c} }
       {|| \perp^{h_c/d}_{h_c}\mathbf{u}_{h_c/d}
         - \mathbf{u}_\textrm{exact} ||_{h_c} }\, ,
  \label{exact_conv_ratio}
\end{align}
where we decrease grid spacing by a factor of~$d$ when increasing
resolution. $\mathcal{C}_\textrm{exact}$ is cheaper numerically
than~$\mathcal{C}_\textrm{self}$ since only two different resolutions
are required to build it, and again the exact solution is understood
to be evaluated on the grid itself. It is possible for a scheme to be
self-convergent but fail to be convergent, for example if one were to
implement the wrong field equations in error. Therefore one would like
to compare the numerical solution to an exact solution wherever
(rarely) possible. To calculate~$\mathcal{C}_{\textrm{exact}}$ we
compute the discretized approximation to a suitable continuum norm at
two resolutions, one twice the other. Each are computed on the
naturally associated grid. We then take the ratio of the two at shared
timesteps, corresponding to those of the coarse grid~$h_c$. In our
setup~$\mathbf{u}_\textrm{exact} = \mathbf{0}$ and~$d=2$, hence
\begin{align}
  \mathcal{C}_\textrm{exact} = \log_2
  \frac{|| \mathbf{u}_{h_c} ||_{h_c} }
  {|| \perp^{h_c/2}_{h_c}\mathbf{u}_{h_c/2} ||_{h_c} } 
  \,,\label{exact_conv_ratio_log2}
\end{align}
which again equals two for perfect convergence. As previously
mentioned appropriate scaling of the random noise amplitude is crucial
and is determined by the norm in which we wish to test convergence. To
realize the proper scaling in our setup, let us consider the exact
convergence ratio~\eqref{exact_conv_ratio_log2} and denote
as~$A_{h_c}$ and~$A_{h_c/2}$ the amplitude of the random noise for
simulations with resolution~$h_c$ and~$h_c/2$ respectively
\begin{align*}
  \mathcal{C}_\textrm{exact} = \log_2
  \frac{||\mathbf{u}_{h_c}||_{h_c}}{||\perp^{h_c/2}_{h_c}
    \mathbf{u}_{h_c/2}||_{h_c}} \sim \log_2
  \frac{O(A_{h_c})}{O(A_{h_c/2})} \, .
\end{align*}
The above suggests that to construct noisy data that converge in the
discretized version of the~$L^2$-norm~\eqref{discrete_L2} for our
second order accurate numerical scheme, we need to drop the amplitude
of the random noise by a quarter every time we double resolution. For
convergence tests in the lopsided norm the scaling factor is
different, due to the~$D_z \phi$ term that appears in the discretized
version of the lopsided norm~\eqref{discrete_lopsided}. By replacing
the~$L^2$ with the lopsided norm in~\eqref{exact_conv_ratio_log2} we
get
\begin{align*}
  \mathcal{C}_{\textrm{exact}} =
  \log_2
  \frac{||\mathbf{u}_{h_c}||_{q (h_c)}}{||
  \perp^{h_c/2}_{h_c}\mathbf{u}_{h_c/2}||_{q (h_c)}}
  \sim \log_2 \frac{O(A_{h_c})}{2 \, O(A_{h_c/2})}\, ,
\end{align*}
where now the norm estimate is dominated by the~$D_z \phi$
term. Hence, to construct noisy data that converge in the lopsided
norm for our second order accurate numerical scheme, we need to
multiply the amplitude of the random noise with a factor of one eighth
every time we double resolution. This discussion would be more
complicated if we were using either pseudospectral approximation or
some hybrid scheme, which is why we focus exclusively on a
straightforward finite differencing setup.

The results for norm convergence tests with appropriately scaled noisy
data for the~$L^2$-norm, for both SH and WH models, are collected in
the right column of Fig.~\ref{fig:L2_norms_all}. As illustrated there,
the inhomogeneous SH model still exhibits convergence since with
increasing resolution the exact convergence ratio tends closer to the
desired value of two at all times of the evolution. On the contrary,
the homogeneous WH model does not converge, and it becomes clear that
with increasing resolution the exact convergence ratio of this model
moves further away from two at all times.

\begin{figure*}[t]
  \includegraphics[width=1.\textwidth]{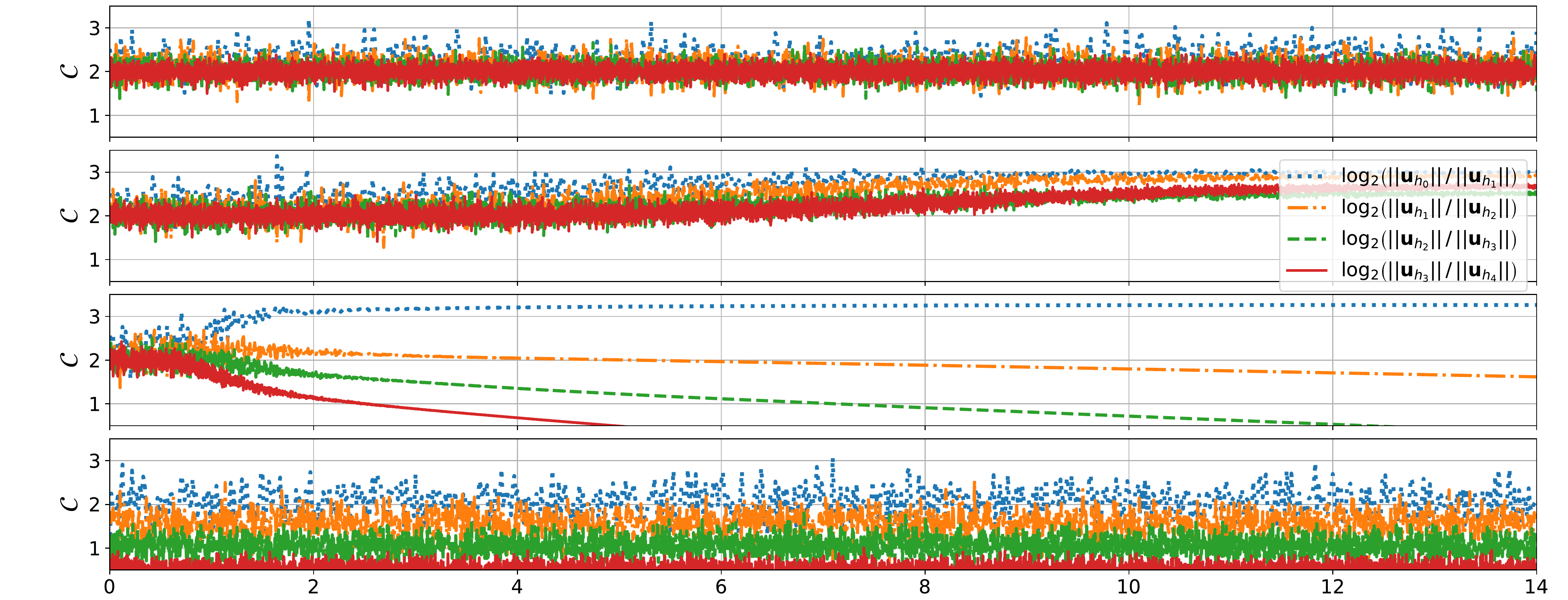}
  \centering{
    \caption{The exact convergence ratio in the lopsided
      norm~\eqref{discrete_lopsided} for the different WH models. From
      top to bottom we plot the homogeneous WH model, then the
      inhomogeneous adjustments, in order~$B_1,B_2$ and~$B_3$. Overall
      we conclude that the homogeneous model and~$B_1$ models are
      converging in the limit of infinite resolution, with the others
      failing to do so. Of these, all but the third panel, with
      source~$B_2$, agree with our expectation from continuum
      considerations. In this one case our method appears to have an
      honest numerical instability, which could be understood properly
      by careful consideration of the
      scheme.\label{fig:lopsided_norms_all}}}
\end{figure*}

To appreciate {\it intuitively} why noisy data allow us to diagnose a
lack of strong hyperbolicity, consider the systems in frequency space
as in subsection~\ref{Subsection:algebraic_characterization}, which we
may think of as momentum space. In practical terms,
Eqn.~\eqref{homogeneous_WH_norm} states that the homogeneous WH model
does not satisfy condition~\eqref{wp_inequality}, and so high
frequency modes can grow arbitrarily fast. Considering smooth data
however, predominantly low frequency modes are excited, and so using
our discretized approximation the violation of
inequality~\eqref{wp_inequality} is not visible at the limited
resolutions we employ. Noisy data on the contrary excite substantially
both high and low frequency modes, with the former crucial to
illustrate the violation.

We also perform convergence tests in the lopsided
norm~\eqref{discrete_lopsided} to examine the behavior of the
different WH models. As in the previous setup, in these tests we
monitor the exact convergence ratio as a function of the simulation
time. As illustrated in Fig.~\ref{fig:lopsided_norms_all}, our
expectations from
subsection~\ref{Subsection:algebraic_characterization} for the
homogeneous model are verified. The homogeneous WH model converges at
all times in the lopsided norm, provided of course that the given data
are restricted to converge at second order to the trivial solution in
the same norm. As also expected, the inhomogeneous case with~$B_3$
fails to converge whatsoever during the evolution, exhibiting behavior
similar to the homogeneous WH model in the~$L^2$-norm tests. In fact,
in this test the exact convergence ratio diverges further from two
with increasing resolution and at earlier times. The discussion for
the inhomogeneous WH models with sources~$B_1$ and~$B_2$ is more
subtle. Both cases initially exhibit convergence, with the~$B_1$ case
maintaining this behavior for longer. The difference lies in their
late time behavior and their trend with increasing resolution.  In
particular, the~$B_1$ case converges for longer with increasing
resolution whereas~$B_2$ does the opposite. At late times in the~$B_1$
case~$\mathcal{C}_\textrm{exact}$ reaches a plateau that converges to
two with increasing resolution, which is not true with
sources~$B_2$. Thus our numerical evidence seems to indicate that
the~$B_1$ inhomogeneous WH model converges in the lopsided norm, but
to disagree with the theoretical expectation at the continuum that
the~$B_2$ case does so too. This is not in contradiction with our
earlier calculations however, because, as a careful examination of the
approximation could reveal, purely algorithmic shortcomings may render
a scheme nonconvergent.

\section{Conclusions} \label{Section:conclusions}

Single-null formulations of GR are popular for applications in
numerical relativity in various settings. In asymptotically flat
spacetimes they are used with compactified coordinates to compute
gravitational waveforms at future-null infinity. In asymptotically AdS
spacetimes they are used to compute {\it in} from the timelike
conformal boundary. But relatively little attention has been paid to
well-posedness of the resulting PDE problems, which serves as an
obstacle to the construction of rigorous error-estimates from
computational work. Presently, therefore, we have examined two popular
formulations, the Bondi-Sachs and affine-null systems, and performed
numerical tests for toy models that illustrate the relevance of our
findings. We found in a free-evolution analysis that, due to the
non-diagonalizability of their angular principal part matrices, both
are only weakly hyperbolic.

Our analysis employed a first order reduction, but was sufficiently
general to rule out the existence of any other reduction (at least
within a large class) that is strongly hyperbolic. We showed also that
the degeneracy can not be avoided by a change of frame. Text book
results on these systems then show that they are ill-posed in
the~$L^2$-norm or its obvious derivatives. Considering model problems
of a similar structure we saw that the same result naturally carries
over to the CIBVP. In the latter case care is needed not to confuse
the usual degeneracy of the norms that appear naturally in
characteristic problems with high-frequency blow-up of solutions. It
follows that a numerical approximation cannot converge to the exact
solution of these PDE problems in any discrete approximation
to~$L^2$. We demonstrated this shortcoming numerically using our
models and adapting the well-known robust-stability test. Spotting
this shortcoming in practice is subtle because smooth data may, and
often do, give misleading results.

Although our weakly hyperbolic toy model is ill-posed in~$L^2$, it may
be well-posed in a lopsided norm in which the angular derivative of
some specific components of the state vector are included. Thus in
such a case one must be able to control the size of not only the
elements of the state vector in the given data, but also some of their
derivatives. This weaker notion of well-posedness is sensitive to the
presence lower order source terms. For example, our weakly hyperbolic
model is well-posed in a (specific) lopsided norm if it is
homogeneous, or inhomogeneous with sources that respect the nested
structure of the equations intrinsic to the characteristic
hypersurfaces. If this nested structure is broken by the source terms,
it becomes ill-posed in any sense. Again using random noise for
initial data, our numerical experiments are consistent with this
analytic result. There is one case in which convergence is not
apparent in our approximation, despite the well-posedness of the
continuum equations in the lopsided norm. This is our only example of
a {\it pure} numerical instability, and is important as it highlights
the fact that for weakly hyperbolic systems numerical methods are not
well-developed, and are not guaranteed to converge, even when using
lopsided norms.

Bringing our attention back to the characteristic initial boundary
value problem for GR, which covers both CCE and applications in
numerical holography, it is clear that the two formulations we
considered will be ill-posed in~$L^2$. It is not clear however, in
general, if they will admit estimates in suitable lopsided norms. But
since the field equations {\it do} have a nested-structure, and our
weakly hyperbolic model problem turned out to admit estimates in
lopsided norms whenever this structure was present, there is reason to
be hopeful. On the other hand, given this uncertainty, and the fact
that numerical approximation to weakly hyperbolic systems (using
lopsided norms) is poorly understood, it is desirable to obtain and
adopt strongly or ideally symmetric hyperbolic alternatives. These
could be sought out by changing gauge directly, or by the use of a
dual-foliation formulation as suggested in~\cite{Hil15}. Perhaps a
simpler option would be to pay the price of evolving curvature
quantities as variables. Several such formulations are known to be
symmetric hyperbolic in a double-null
gauge~\cite{CabChrTag14,HilValZha19,HilValZha20} and could be adjusted
appropriately.

A true {\it principle} solution to wave-extraction would be a robust
scheme for CCM, the other main alternative being the use of
compactified hyperboloidal slices, a topic also under active research
for full
GR~\cite{BarSarBuc11,Zen10,VanHusHil14,Van15,DouFra16,HilHarBug16,
  VanHus17,GasHil18,GasGauHil19,BeyFraHen20}. To understand the
consequences of our findings for CCM we considered a model in which
the IBVP is solved for a symmetric hyperbolic system, and the
solutions are then glued through boundary conditions to those of a
weakly hyperbolic system accepting estimates in lopsided norms. The
former of these two sets of equations is viewed as a model for the
formulation used in the strong-field region, the latter for a
single-null formulation used on the outer characteristic domain. With
this setup, we found that the fundamental incompatibility of the norms
naturally associated with the two domains prohibits their combined use
in building estimates. But if the weakly hyperbolic system were made
symmetric hyperbolic progress could be made. A less appealing
possibility would be to demonstrate that the formulation in the Cauchy
domain, or some suitable replacement, admits estimates in a lopsided
norm compatible with that of the characteristic region. Since this
relies on very special structure in the field equations, the outlook
for a complete proof of well-posedness of CCM using existing
Bondi-like gauges is, unfortunately, not rosy.

Our results signpost a number of paths to follow. First and foremost,
we need to recover our numerical results for toy models for full
GR. Beyond that, we seek a well-posed setup for the CIBVP that can be
used in numerical applications with minimum change to existing code.
For the latter it will be useful to perform a pure gauge analysis
along the lines of~\cite{KhoNov02,HilRic13} to establish whether or
not the blame for the degeneracy can be unambiguously laid on the
coordinate choice, or if the specific construction of the formulations
we discussed have some influence. Work in both directions is ongoing.

\acknowledgments

We are grateful to Thomas Baumgarte, Nigel Bishop, Carsten Gundlach,
Luis Lehner and Denis Pollney for helpful discussions and/or comments
on the manuscript. We also thank Mikel S\'anchez for feedback on our
Julia scripts. The work was partially supported by the FCT (Portugal)
IF Program~IF/00577/2015, IF/00729/2015, PTDC/MAT-APL/30043/2017 and
Project~No.~UIDB/00099/2020. TG acknowledges financial support
provided by FCT/Portugal Grant No. PD/BD/135425/2017 in the framework
of the Doctoral Programme IDPASC-Portugal. The authors would like to
acknowledge networking support by the GWverse COST Action CA16104,
``Black holes, gravitational waves and fundamental physics''.

\bibliography{refs}

\end{document}